\documentclass[a4paper,11pt]{article}

\usepackage{jheppub} 

\usepackage[T1]{fontenc} 

\title{\boldmath Quark masses in Higgs production with a jet veto}

\author[a]{Andrea Banfi,}
\author[b,c,d]{Pier Francesco Monni,}
\author[c]{Giulia Zanderighi}

\affiliation[a]{Department of Physics and Astronomy, University of Sussex,\\Sussex House, Brighton, BN1 9RH, U.K.}
\affiliation[b]{Institut f\"ur Theoretische Physik, Universit\"at
  Z\"urich, \\Winterthurerstrasse 190, CH-8057, Z\"urich, Switzerland}
\affiliation[c]{Rudolf Peierls Centre for Thoeretical
  Physics,University of Oxford,\\1 Keble Road, Oxford OX1 3NP, U.K.}
  \affiliation[d]{
  Institute for Particle Physics Phenomenology, University of Durham,
  \\ South Road, Durham DH1 3LE, U.K.}

\emailAdd{a.banfi@sussex.ac.uk}
\emailAdd{pfmonni@physik.uzh.ch}
\emailAdd{g.zanderighi1@physics.ox.ac.uk}

\abstract{We study the impact of finite mass effects due to top and
  bottom loops in the jet-veto distribution for Higgs production. We
  discuss the appearance of non-factorizing logarithms in the region
  $\ptjv \gtrsim m_b$.  We study their numerical impact and argue that
  these terms can be treated as a finite remainder. We therefore
  detail our prescription for resumming the jet-vetoed cross section
  and for assessing its uncertainty in the presence of finite mass
  effects. Resummation for the jet-veto, including mass effects, has
  been implemented in the public code {\tt JetVHeto}.}

\preprint{OUTP-13-16P, ZU-TH 17/13}

\newcommand{\as}{\alpha_s}

\newcommand{\TeV}{\;\mathrm{TeV}}
\newcommand{\ptjv}{p_{\rm t,veto}}
\newcommand{\cF}{{\cal F}}
\newcommand{\cO}[1]{{\cal O}\left(#1\right)}
\newcommand{\NNLL}{\text{NNLL}}
\newcommand{\eff}{\epsilon}

\begin{document} 
\maketitle
\flushbottom

\section{Introduction}

A new particle with mass $m_H \sim 125$ GeV was discovered last year
at the LHC~\cite{Aad:2012tfa,Chatrchyan:2012ufa}.
After just one year from its discovery, precision measurements of the
characteristics of this particle have already been carried out: its
mass is known with better than $1$ GeV precision, and all properties
studied so far are compatible with the particle being the Higgs boson
of the Standard Model (SM).
This marks a remarkable success of the experiment. However, contrary
to expectations, no evidence for New Physics has been found yet at the
LHC.  Since New Physics is elusive, it is natural to focus on the
newly discovered particle and to perform as accurate as possible
phenomenological studies on it.
In order to fully establish the nature of the new particle as the SM
Higgs boson, even more precise measurements of its properties, in
particular its couplings, are essential.  Establishing a significant
departure from the simple SM-like pattern could rule out the
possibility that the new particle is the plain SM Higgs boson, and
could be a first manifestation of New Physics.

In fact, the new particle lies in a sweet spot where many decay modes
have an appreciable branching fraction, leading to a very rich
phenomenology. Second only to the branching to $b\bar b$, which
however is very difficult experimentally because of the large $gg\to b
\bar b$ background, is the branching to $WW$. In this channel, a
dominant, reducible background comes from top-pair production, where
each top decays to a $W$-boson and a $b$-quark that generally gives
rise to a jet.
On the other hand, most Higgs-boson events involve either none or one
jet, since higher multiplicity events are suppressed by powers of the
strong coupling constant, and hence more rare.
It is natural then to impose a veto on additional jets so as to
considerably reduce the dominant top background, while reducing only
modestly the signal.
Still, it is then essential to understand how much these tight
kinematical cuts reduce the signal cross-section, and what are the
uncertainties associated to efficiencies and cross-sections in the
presence of jet-vetoes.

Perturbative predictions involve, for each power of the coupling
constant, double logarithms of the jet-veto scale $\ptjv$ and of the
scale of the hard scattering $m_H$. ATLAS and CMS use jet-veto scales
of the order of $25$-$30$ GeV, so that in order to make accurate
predictions it becomes important to account for these logarithms to
all orders in perturbation theory. In recent years, a lot of progress
has been made both in understanding and computing resummed predictions
for the jet-veto cross section. In ref.~\cite{Banfi:2012yh}, it was
pointed out that the jet-veto is within the scope of the resummation
program \texttt{CAESAR}~\cite{Banfi:2004yd} and a next-to-leading
logarithmic (NLL) resummation matched to next-to-next-to-leading order
(NNLO) results was presented.
Subsequently, a full NNLL+NNLO resummation was computed
in~\cite{Becher:2012qa,Banfi:2012jm} and very recently also
in~\cite{Stewart:2013faa,Becher:2013xia}, including partial N$^3$LL
terms.
However, a full N$^3$LL resummation requires still some work, among
which the exact calculation of the four-loop cusp anomalous dimension
(rather than its Pad\'e approximation) and the exact coefficient of
the relative $\alpha_s^3 L$ term.

Residual uncertainties for jet-veto efficiencies for a $25$-$30$ GeV
jet-veto are of the order of $\pm 10\%$. These uncertainties propagate
into the jet-veto cross-section, giving rise to uncertainties of the
order of $\pm$11-12\% (for 8 TeV collisions, using the anti-$k_t$
algorithm~\cite{Cacciari:2008gp} with $R=0.5$ with $\ptjv=25$-$30$
GeV). This should be compared with a $\pm$15-20\% uncertainty in
fixed-order NNLO predictions.
While the resummation is successful in reducing the uncertainties in
this region, one needs to bear in mind that all these resummed
predictions use the large-$m_t$ effective theory, where the Higgs
couples directly to gluons via an effective vertex, and bottom quark
loops are completely neglected. The above uncertainties do not account
for the error associated with the large-$m_t$ approximation.  Given
the high accuracy reached by these resummed predictions, it becomes
then important to assess the impact of finite-mass effects on
jet-vetoed cross-sections and the associated theoretical error.

The impact of finite mass corrections in Higgs cross-sections or Higgs
transverse momentum distributions has been studied already in quite
some detail at fixed
order~\cite{Spira:1995rr,Spira:1997dg,Harlander:2005rq,Anastasiou:2006hc,Aglietti:2006tp,Bonciani:2007ex,Harlander:2012hf}.
Exact finite-mass effects for Higgs production accompanied by one jet
have been also implemented at LO in the code {\tt MCFM}~\cite{MCFM}.
In Ref.~\cite{Anastasiou:2009kn}, {\tt HPRO}, a NLO Monte-Carlo for
Higgs production via gluon fusion with finite heavy-quark masses was
presented. In~\cite{Harlander:2012pb} the program {\tt SusHi} is
presented, that contains NLO QCD contributions from the third family
of quarks and squarks, NNLO corrections due to top-quarks, approximate
NNLO corrections due to stops, and electro-weak effects. The
contribution of initial-state bottom quarks has been computed at NNLO
in refs.~\cite{Harlander:2003ai, Harlander:2010cz, Harlander:2011fx, Buehler:2012cu} and it was found to be small.
Furthermore, finite quark-mass effects are included in Monte Carlo
generators like {\tt Herwig}~\cite{Corcella:2000bw,Corcella:2002jc}
and {\tt Pythia}~\cite{Sjostrand:2006za} and have recently been
implemented also in the NLO Monte Carlo event generators {\tt
  POWHEG}~\cite{Bagnaschi:2011tu} and {\tt MC@NLO}~\cite{mcatnlo}.

In general, for inclusive cross-sections one finds that bottom-mass
effects are important and opposite in sign to the top-mass effects. In
fact, while the pure top contribution gives rise at NLO to about a
$6\%$ correction to the inclusive cross-section, once both top and
bottom effects are included the correction with respect to the heavy
top limit is about $-2\%$.
Both top and bottom mass effects distort the Higgs transverse momentum
spectrum considerably, and therefore cannot be included as a
$K$-factor. In particular at large transverse momenta corrections
become large and negative.
Furthermore, while {\tt POWHEG} and {\tt MC@NLO} agree well with each
other when only exact top-quark mass effects are accounted for, when
also $b$-mass effects are included, the two approaches give
considerably different predictions for \mbox{$p_{t,H} \lesssim 50$
  GeV}, i.e. in the region where resummation effects are
important. This is not surprising since in the {\tt POWHEG} approach
higher order terms in the Sudakov are exponentiated, while in {\tt
  MC@NLO} they are not. If these higher order effects are important,
the two methods will give numerically different answers.

An analytic resummation for the Higgs transverse momentum distribution
including mass effects have been considered recently at NLL+NLO in
ref.~\cite{Mantler:2012bj}. Finite top- and bottom-quark masses up to
${\cal O}(\alpha_s^3)$ have also been included in the NNLL+NNLO
transverse momentum spectrum of the Higgs
boson~\cite{Grazzini:2013mca}. This reference also discusses
difficulties arising in the case of bottom quarks, i.e. when the
heavy-quark mass is much smaller than the Higgs mass. In fact, in this
case the two-scale problem (the two scales being $p_{t,H}$ and $m_H$)
that is treated using standard resummation techniques becomes a
three-scale problem (the third scale being the bottom mass $m_b$). The
approach of ref.~\cite{Grazzini:2013mca} is based on the observation
that collinear factorization is valid only when $p_{t,H} \lesssim
m_b$. Hence, the authors of ref.~\cite{Grazzini:2013mca} introduce a
second resummation scale which controls the impact of the resummation
of bottom quark contributions. They set this scale to $m_b$ hence
limiting the resummation only to the region $p_{t,H} \lesssim m_b$.

In this paper, we study the ``factorization breaking terms'' which
appear for $p_{t} \gtrsim m_b$ further and argue that, in our
approach, it is not necessary to switch off the resummation in this
region. We also show that these terms can be safely treated as any
other non-factorizing finite remainder in our resummation formalism.

The paper is organized as follows. In Sec.~\ref{sec:kin} we describe
the kinematics of the problem and introduce the jet-veto
observable. In Sec.~\ref{sec:fact-soft-sing} we discuss the
(non)-factorizing terms appearing when considering soft and collinear
limits of the Higgs plus one parton matrix elements. In
Sec.~\ref{sec:facandres} we assess the numerical impact of these
terms, discuss in detail the prescription we use to treat them and
present our resummation formula including finite-mass effects.
In particular, we discuss whether it is appropriate, and what is the
impact of changing the resummation scale for the bottom induced
contribution and we add an additional logR-type matching scheme, whose
purpose is to explore the impact of exponentiating, or not,
non-factorizing terms whose higher order structure is unknown.
In Sec.~\ref{sec:pheno} we present phenomenological results for the
LHC with realistic cuts as applied by ATLAS and CMS.  We pay
particular attention to quantifying the impact of top and bottom mass
effects by comparing with the large-$m_t$ approximation and to
assessing theory uncertainties. We conclude in
Sec.~\ref{sec:conclu}. Appendix~\ref{sec:app-real} collects results
for real emission matrix element squared, Appendix
~\ref{sec:matchingschemes} collects our standard matching formulae,
and Appendix~\ref{sec:modascheme} presents an additional matching
scheme.

\section{Kinematics and cross sections}
\label{sec:kin}

We consider the production of a Higgs boson accompanied by an
arbitrary number of extra QCD partons
\begin{equation}
p_1 \,p_2 \to p_H\, p_3,\dots,p_n\,.
\end{equation}
A jet-veto condition is imposed by clustering the final state partons
into jets using an infrared and collinear safe jet-algorithm, and
requiring that each event has no jets with transverse momentum above
$\ptjv$. The jet-veto efficiency $\epsilon(\ptjv)$ is defined as the
fraction of events that pass the jet-veto condition:
\begin{equation}
  \epsilon(\ptjv)= \frac{\Sigma(\ptjv)}{\sigma_{\rm tot}}\,.
\label{eq:jetvdef}
\end{equation}
Here $\Sigma(\ptjv)$ is the so-called zero-jet cross section, defined
as
\begin{equation}
  \Sigma(\ptjv) = \sum_n \int d\Phi_n \frac{d\sigma_n}{d\Phi_n} \Theta(\ptjv - p_{\rm t, jet})\,,
\end{equation}
where $p_{\rm t, jet}$ is the transverse momentum of the hardest jet,
$d\sigma_n$ is the cross section for producing a Higgs and $n\!-\!2$
extra partons, and $d\Phi_n$ is the corresponding phase
space. Moreover, $\sigma_{\rm tot}$ is the total Higgs production
cross section.  Both $\sigma_{\rm tot}$ and $\Sigma(\ptjv)$ can be
expanded in powers of $\alpha_s$
\begin{subequations}
  \begin{align}
    \label{eq:Sigma-expansion}
    \Sigma(\ptjv) &= \Sigma_{0}(\ptjv) + \Sigma_{1}(\ptjv) + \Sigma_{2}(\ptjv) + \ldots \,,\\
    \eff(\ptjv) &= \eff_{0}(\ptjv) + \eff_{1}(\ptjv) + \eff_{2}(\ptjv) + \ldots \,,\\
    \sigma_{\rm tot} &= \sigma_{0} + \sigma_{1} + \sigma_{2} + \ldots \,.
  \end{align}
\end{subequations}
At the lowest order, $\alpha_s^2$, $ \Sigma_{0}(\ptjv)$ is just the LO
cross section $\sigma_0$. At higher orders, we obtain the zero-jet
cross section from the differential distribution in $p_{\rm t, jet}$
as follows
\begin{equation}
  \label{eq:sigmabar}
  \Sigma_i(\ptjv) = \sigma_i+\bar\Sigma_i(\ptjv)\,,\qquad \bar
  \Sigma_i(\ptjv) = -\int_{\ptjv}^{\infty} \!\!\!\! dp_t \frac{d\Sigma_i}{dp_t}\,.
\end{equation}

Starting from NLO, each $\Sigma_i(\ptjv)$ contains large logarithmic
corrections up to $\alpha_s^{2+i} L^{2i}$, with $L=\ln(m_H/\ptjv)$. In
ref.~\cite{Banfi:2012jm}, such logarithmic contributions have been
resummed in the context of the \texttt{CAESAR}
approach~\cite{Banfi:2004yd} up to NNLL accuracy, i.e.\ accounting for
all logarithmically enhanced terms down to $\alpha_s^n L^{n-1}$ in
$\ln(\Sigma/\sigma_0)$. The basis of the resummation in
ref.~\cite{Banfi:2012jm} is the factorization of soft and collinear
singularities from the Born matrix element. At tree level, the Higgs
is produced via gluon fusion through a top loop. If the top-quark mass
$m_t$ is much larger than all other scales in the problem, {\it i.e.}
the large-$m_t$ limit, factorization of soft and collinear
singularities is valid since soft or collinear partons can never
resolve the top loop.
However, when the transverse momentum of the emitted gluons is larger
than the mass of the virtual quarks, new logarithms of the ratio of
the two scales appear in the perturbative expansion.  Such logarithms
are not proportional to the Born cross section, and in the following
we will refer to them as non-factorizing terms. In the case of light
quarks, {\it i.e.} bottom quarks, these logarithmic terms can
potentially have an important impact in the region of phenomenological
interest.\footnote{The Yukawa coupling and explicit mass dependence of
  quark-loops suppress the impact of yet lighter quarks.}

The appearance of these new logarithms can be observed already at the
Born level for the case of Higgs production with no extra partons. Let
us first consider a single quark loop, and let $m_H$ and $m$ denote
the Higgs and quark mass, respectively. The amplitude squared for the
process is given by
\begin{equation}
  \label{eq:M2higgs}
  |M_B|^2 = (N_c^2-1)\frac{y^2 \alpha_s^2}{64 \pi^2} \left(|M_{++}|^2+|M_{--}|^2\right)\,,
\end{equation}
where $y = g_w m /(2M_W)$ is the quark Yukawa coupling, and $M_{++}$,
$M_{--}$ are the only two non-vanishing helicity amplitudes, since the
helicity of the incoming gluons has to be the same due to angular
momentum conservation. Furthermore thanks to parity conservation
$M_{++} = M_{--}$.

The amplitude $M_{++}$ (taken from ref.~\cite{Ellis:2011cr}
Eq. (9.34)) is given by
\begin{equation}
  \label{eq:M++}
  M_{++} = 8m\left[\left(1-\frac{4 m^2}{m_H^2}\right)\frac{1}{2} m_H^2 C(m_H^2)-1\right]\,,
\end{equation}
where $C(m_H^2)$ is the scalar triangle integral in
eq.~\eqref{eq:triangle}. If $m^2\gg m_H^2$ we recover the well-known
large-$m$ limit, 
\begin{equation}
  \label{eq:M2higgs-infinity}
  |M_B^\infty|^2 =  (N_c^2-1)\frac{G_F \, m_H^4}{9\sqrt{2}}\frac{\alpha_s^2}{\pi^2}\,. 
\end{equation}

If $m^2\ll m_H^2$ the triangle integral reads 
\begin{equation}
  \label{eq:triangle-mllmH}
  m_H^2 C(m_H^2) =
  \frac{1}{2}\ln^2\left(-\frac{m_H^2}{m^2}\right)+\mathcal{O} \left(\frac{m^2}{m_H^2}\right)\,,
\end{equation}
which gives
\begin{equation}
  \label{eq:M++mllmH}
  M_{++} \simeq 8m\left[\frac{1}{2} m_H^2 C(m_H^2)-1\right]  \simeq 2m \ln^2\left(-\frac{m_H^2}{m^2}\right)\,.
\end{equation}
The physical origin of the result above can be understood by examining
the structure of the quark loop.  First of all, the Higgs-quark vertex
does not conserve helicity, therefore the amplitude has to be
proportional to the quark mass $m$.  When the loop momentum probes the
kinematical region between the two relevant scales (the quark mass $m$
and the Higgs mass $m_H$), the loop integral exhibits the usual
(soft-collinear) double logarithmic behaviour.

If one extra gluon with transverse momentum $p_t\geq m $ is emitted
off the fermion line running in the loop, then one expects to observe
double logarithms of the type $\ln^2(p_t^2/m^2)$.  For the jet-veto
cuts applied by ATLAS and CMS, $\ptjv = 25$-$30$ GeV, $\ptjv/m_b \sim
m_H/\ptjv$, therefore these logarithms can be as large as the
logarithms $\ln(m_H/\ptjv)$ that we want to resum. Furthermore these
logarithms are not proportional to the Born amplitude in
eq.~\eqref{eq:M++mllmH}, meaning that soft radiation does not
factorize in the regime $m^2 \ll p_t^2$.
Since this fact questions the basis of our resummation approach, we
devote the next section to the calculation of these non-factorizing
corrections to lowest order in perturbation theory.
 
\section{Factorization of soft and collinear singularities}
\label{sec:fact-soft-sing}

We consider here the amplitude for Higgs production in association
with one extra parton
\begin{equation}
  \label{eq:H+1parton}
  p_1 p_2 \to p_H p_3\,.
\end{equation}
At lowest order in perturbation theory this process was computed
in~\cite{Ellis:1987xu,Baur:1989cm}. It has three contributing
subprocesses, $gg \to H g$, $qg \to H q$, and $g\bar q \to H \bar q$.
These subprocesses proceed through gluon fusion via a quark loop.  The
case when the mass of the virtual quark is comparable to or heavier
than the Higgs mass is well-known.
We wish to investigate the behaviour of the amplitudes in presence of
a light quark of mass $m\ll m_H$ when the extra parton of transverse
momentum $p_t$ is either soft or collinear.  We are interested in both
cases $m\ll p_t \ll m_H$ and $p_t \ll m \ll m_H$.

\subsection{Soft limit}
\label{sec:soft-limit}

We consider first the case in which the emitted parton $p_3$ is
soft. As expected, and by direct inspection of the amplitudes in
Appendix~\ref{sec:app-real}, only the subprocess $gg \to H g$ exhibits
a soft singularity.
It is useful to split the amplitude into different contributions
according to the helicity configurations $\lambda_1, \lambda_2,
\lambda_3$ of the involved gluons $p_1,p_2$ and $p_3$, with
$\lambda_i=\pm$. Relations among the various helicity amplitudes
$M^{\lambda_1 \lambda_2 \lambda_3}_{gg\to Hg}$ can be found in
Appendix~\ref{sec:app-real}, Eqs.~\eqref{eq:M-gluons}
and~\eqref{eq:M-gluonsP}.

Due to angular momentum conservation, in the soft limit, only two of
the four independent amplitudes are non-zero, namely $M_{gg\to
  Hg}^{+++}$ and $M_{gg\to Hg}^{++-}$ (see eqs.~(\ref{eq:M+++}) and
(\ref{eq:M++-})).
In this limit, the invariants $s,t,u$ in eq.~\eqref{eq:invariants},
and $s_1,u_1,t_1$ in eq.~\eqref{eq:new-invariants} reduce to
\begin{equation}
  \label{eq:invariants-soft}
  s \to m_H^2\,,\quad t, u \to 0\,,\quad s_1\to 0\,,\quad t_1,u_1\to -m_H^2\,.
\end{equation}
Furthermore, in the soft limit, we keep only the leading terms
proportional to $1/(ut)=1/(m_H^2 p_t^2)$ and drop all terms
proportional to $m^2$. This gives
\begin{multline}
  \label{eq:M+++soft}
    M_{gg\to Hg}^{+++}\simeq-\frac{64 \,m \,y\, \Delta}{ut}\left\{1-\frac{1}{2} m_H^2 C(m_H^2)\right.\\
    \left.+\frac{1}{4}\left[tC(t)+uC(u)-\frac{1}{2}\left[st D(s,t)+usD(u,s)-utD(u,t)\right]\right]\right\}\,,
\end{multline}
and
\begin{multline}
  \label{eq:M++-soft}
    M_{gg\to Hg}^{++-}
\simeq \frac{64\, m\, y\, \Delta}{ut}\left\{1-\frac{1}{2} m_H^2 C(m_H^2)\right.\\
\left.+\frac{1}{4}\left[tC(t)+uC(u)-\frac{1}{2}\left[st D(s,t)+usD(u,s)+utD(u,t)\right]\right]\right\}\,,
\end{multline}
where $\Delta=\sqrt{(stu)/8}$.  Notice that both amplitudes contain a
factorizing term (first line) that is proportional to the Born
amplitude eq.~(\ref{eq:M++}), and an additional term, which we call a
non-factorizing correction (second line).  If the latter term does not
vanish in the limit $p_t \to 0$ then the factorization of soft
radiation is violated.  Its expression depends on the behaviour of the
scalar integrals $C$ and $D$, which is different in the two regimes,
$p_t^2 \ll m^2 \ll m_H^2$ and $m^2 \ll p_t^2 \ll m_H^2$.
\begin{enumerate}
\item $p_t^2 \ll m^2 \ll m_H^2$. In this limit, the contributions from
  all $C$ and $D$ integrals vanish (see eqs.~\eqref{eq:triangle} and
  \eqref{eq:box}), except for $m_H^2 C(m_H^2)$. As expected, we have
  factorization of the soft matrix element from the Born amplitude
  $M_{++}$ in eq.~\eqref{eq:M++}
  \begin{equation}
    \label{eq:M-ptllm}
    M_{gg\to Hg}^{+++} =- M_{gg\to Hg}^{++-} \simeq \frac{8 y
    \Delta}{ut}M_{++}\,.
  \end{equation}
\item $m^2 \ll p_t^2 \ll m_H^2$. In this case any triangle integral
  has the behaviour
  \begin{equation}
    \label{eq:triangle-mllpt}
   q^2 C(q^2) \simeq \frac{1}{2} \ln^2\left(\frac{m^2}{-(q^2+ i \epsilon)}\right)\,,
  \end{equation}
and the box integrals reduce to 
  \begin{equation}
    \label{eq:box-mllpt}
    \begin{split}
    st D(s,t)& \simeq \ln^2\left(-\frac{m^2}{t}\right)\,,\quad 
su D(s,u) \simeq \ln^2\left(-\frac{m^2}{u}\right)\,,\\
 ut D(u,t)&\simeq 
 \ln^2\left(-\frac{m^2}{t}\right)+ \ln^2\left(-\frac{m^2}{u}\right)- \ln^2\left(\frac{m^2}{m_H^2}\right)
+2 \ln\left(\frac{-t}{m_H^2}\right) \ln\left(\frac{-u}{m_H^2}\right)\\
&-2 \,i \pi \left[\ln\left(\frac{m^2 m_H^2}{u t}\right)+2 \ln\left(\frac{m_H^4}{u t}\right)\right]-\frac{2}{3} \pi^2\,.
    \end{split}
  \end{equation}
This gives 
\begin{equation}
  \label{eq:M+++mllpt}
  \begin{split}
M_{gg\to Hg}^{+++} &\simeq-\frac{64\,m y\,\Delta}{ut}\left\{1-\frac{1}{2}
    m_H^2 C(m_H^2)+\frac{1}{8} ut D(u,t)
\right\}\\
  &=\frac{8y\,\Delta}{ut} M_{++}
    -8my\,\Delta \,D(u,t)
    \,,
  \end{split}
\end{equation}
and
\begin{equation}
  \label{eq:M++-mllpt}
  \begin{split}
M_{gg\to Hg}^{++-} &\simeq\frac{64\,m\,y\,\Delta}{ut}\left\{1-\frac{1}{2}
    m_H^2 C(m_H^2)-\frac{1}{8} ut D(u,t)
\right\}\\
  &=-\frac{8\,y\,\Delta}{ut} M_{++}
    -8m\,y\,\Delta \,D(u,t)\,.
  \end{split}
\end{equation}
In this regime, non-factorizing terms survive and give a
logarithmically enhanced contribution, proportional to $(1/p_t)
\ln^2(m^2/p_t^2)$ (see eq.~(\ref{eq:box-mllpt})).
We stress that those non-factorizing terms vanish for $p_t$ below $m$
(see e.g. eq.  (\ref{eq:M-ptllm})), so that in the singular limit
$p_t\to 0$ standard factorization of soft singularities is preserved.
Furthermore, we argue that the non-factorizing terms above have a
moderate impact on the squared amplitude. Indeed, in the amplitude
squared the dominant contribution from $b$ quarks comes from the
interference of bottom and top loops.  For the top-loop contribution,
$p_t^2 \ll m_t^2$, the standard factorization holds. Since
non-factorizing terms have opposite signs (with respect to factorizing
ones) in $M_{gg\to Hg}^{+++}$ and $M_{gg\to Hg}^{++-}$, see
eqs. (\ref{eq:M+++mllpt}) and (\ref{eq:M++-mllpt}), they will cancel
in the interference at this perturbative order. Non-vanishing terms in
the amplitude squared appear in the pure $b$-quark amplitude squared
and are of relative order $(m_b/m_H)^4\ln^4(m_b^2/p_t^2)$ with respect
to the dominant top contribution.
\end{enumerate}

\subsection{Collinear limit}
\label{sec:collinear-limit}

We now consider the limit in which the emitted parton $p_3$ is
collinear to e.g. $p_1$. Obviously, the same conclusions will apply
to the case when $p_3$ is collinear to $p_2$ ($u \leftrightarrow
t$). In this limit, the invariant $u$ vanishes, and, defining $z =
m_H^2/s$, we have $t\to - (1-z) s$ and $u = - p_t^2/(1-z)$.
Correspondingly the auxiliary invariants $s_1$, $t_1$ and $u_1$ become
\begin{equation}
  \label{eq:collinear-limit-s1t1u1}
  s_1 \to -t\,,  \quad t_1\to -s\,,\quad \quad u_1 \to -m_H^2.
\end{equation}
In this case we expect the amplitude squared to exhibit a collinear
singularity $1/u$. For simplicity, we consider the matrix element for
the process $qg \to H q$, where $p_1$ and $p_3$ are the momenta of the
incoming and outgoing quarks, respectively. Taking the collinear limit
of the corresponding matrix element squared (see
eq.~(\ref{eq:Mqg-Mgqbar})), we obtain
\begin{equation}
  \label{eq:Mqg-collinear}
  \sum |M_{gq\to Hq}|^2 \simeq \frac{N_c^2-1}{2} \frac{\alpha_s^3}{\pi}  \frac{1+(1-z)^2}{z}\left( \frac{1}{-z u} \right)|\mathcal A(u,t,s)|^2\,,
\end{equation}
where $\mathcal A(u,t,s)$ is the contribution to the amplitude from a single quark
in the loop, of mass $m$. In the collinear limit $u\to 0$, it is given by:
\begin{equation}
  \label{eq:A-collinear}
  \mathcal A(u,t,s) \simeq y\,m \left(2 + 4m^2C_1(u)+ u C(u) - m_H^2 C(m_H^2)\right)\,.
\end{equation}
As before, we consider two regimes:
\begin{enumerate}
\item $p_t^2 \ll m^2 \ll m_H^2$. As in the soft case, from
  eq.~\eqref{eq:triangle} one sees that the triangle integral $u C(u)$
  vanishes. We then obtain
\begin{equation}
  \label{eq:A-collinear-1}
  \mathcal A(u,t,s) \simeq 2 y\, m \left(1- \frac 12 m_H^2
    C(m_H^2)\right) = -\frac{y}{4} M_{++}\,,
\end{equation}
which is proportional to the Born amplitude in eq.~\eqref{eq:M++}.
\item $m^2 \ll p_t^2 \ll m_H^2$. In this case $uC(u)$ does not vanish,
  and we have
\begin{equation}
\label{eq:A-collinear-2}
\begin{split}
  \mathcal A(u,t,s) & \simeq 2 y\,m \left(1 -\frac 12 m_H^2 C(m_H^2) +\frac 12 u
    C(u)\right) \\& 
\simeq -\frac{y}{4}M_{++} +y\frac{m}{2} \ln^2\left(\frac{m^2}{-u}\right) \,.
\end{split}
\end{equation}
In this regime we do not have a factorization of the Born matrix
element from the collinear singularity. Unlike in the soft case, there
is no other amplitude against which this contribution can cancel,
therefore a non-factorizing correction of order $(m_b/m_H)^2
\ln^2(p_t/m_b)$ remains in the amplitude. The presence of
non-factorizing terms in the collinear limit has been noted also in
ref.~\cite{Grazzini:2013mca}.  Similar contributions appear in the
pure gluonic subprocess.
\end{enumerate}

\section{Factorization and resummation formula}
\label{sec:facandres}

\subsection{Resummation formula}
\label{sec:resumformulae}
In the previous section we have seen that, in the region $m^2 \ll
p_t^2 \ll m_H^2$, the $\mathcal{O}(\alpha_s^3)$ matrix element for
Higgs production with an extra jet is the sum of a factorizing and a
non-factorizing term which vanishes in both the soft and collinear
limits.  The non-factorizing term, not proportional to the Born matrix
element, contains a new class of logarithms $\ln(\ptjv/m_b)$. At
typical veto scales $\ptjv=25$-$30\,\mathrm{GeV}$, these logarithms
can be potentially as large as the factorizing Sudakov logarithms
$\ln(m_H/\ptjv)$ that we wish to resum. However, in the matrix element
squared, their coefficients are suppressed by a factor $(m_b/m_H)^2$
in the collinear limit, and $(m_b/m_H)^4$ in the soft limit. Therefore
it is important to investigate their actual numerical impact, and we
will do so in the next subsection.

Since we do not know the structure of finite quark-mass corrections
beyond ${\cal O}(\alpha_s^3)$, matching schemes need to preserve the
structure of the perturbative series up to this order.  The three
multiplicative matching schemes defined in~\cite{Banfi:2012jm} and
reported in Appendix~\ref{sec:matchingschemes} automatically fulfill
this structure. In these schemes, factorization is assumed to hold
only for the singular terms containing $\ln (m_H/\ptjv)$ which must be
resummed since they become dominant in the low $\ptjv$ region.  The
additional terms (containing $\ln(m_b/\ptjv)$) which vanish in
the limit $\ptjv \to 0$ with $\ptjv > m_b$ are automatically treated
as a regular remainder, which does not multiply the Born cross section.

Starting from this observation, we define our full NNLL resummed cross
section for the jet veto as 
\begin{multline}
  \label{eq:SigmaNNLL-result}
  \Sigma^{(J)}_\text{\NNLL}(\ptjv) =
  \sum_{i,j}\int dx_1 dx_2 \,|{M}_B|^2\delta(x_1 x_2 s - M^2) \times \\ \times \bigg[f_i\!\left(x_1, e^{-L} \mu_F\right)
  f_j\!\left(x_2, e^{-L} \mu_F\right)\left(1+\frac{\alpha_{s}}{2\pi}{\cal H}^{(1)}\right)\\
  + \frac{\alpha_{s}}{2\pi}\frac{1}{1-2\alpha_s \beta_0 L}\sum_{k}\int_{x_1}^1\frac{dz}{z}\bigg(
  C_{ki}^{(1)}(z) f_i\!\left(\frac{x_1}{z}, e^{-L} \mu_F\right)
  \times f_j\!\left(x_2, e^{-L} \mu_F\right) + \{(x_1,i)\,\leftrightarrow\,(x_2,j)\}\bigg)\, \bigg]\times \\
  \times(1 + \cF^{\text{clust}} + \cF^{\text{correl}})
   e^{L g_1(\as L) + g_2(\as L) + \frac{\as}{\pi} g_3(\as L)}\,.
\end{multline}
Here $L=\ln(Q/\ptjv)$, where $Q$ is a resummation scale to be taken of
order $m_H$~\cite{Banfi:2012yh,Banfi:2012jm}.~\footnote{For all plots
  we actually replace $L$ by $\tilde L$ defined in
  Appendix~\ref{sec:matchingschemes}. This ensures that the fixed
  order result is reproduced at high $\ptjv$.}
The coefficient functions $C_{ki}^{(1)}$, and resummation functions
$g_1$, $g_2$ and $g_3$ are the same used in ref.~\cite{Banfi:2012jm}.
The finite-mass effects in eq.~\eqref{eq:SigmaNNLL-result} are
contained both in the Born amplitude $|M_B|^2$ and in the one-loop
coefficient ${\cal H}^{(1)}$ which accounts for the finite part of the
virtual corrections to the Higgs production amplitude at order ${\cal
  O}(\alpha_s^3)$.  The Born amplitude squared is related to its
large-$m_t$ limit $|M_B^{\infty}|^2$ by the following relation
\begin{equation}
| M_B|^2 = | M_B^{\infty}|^2\biggl\lvert\sum_q F_0(\tau_q)\biggr\rvert^2\,,
\end{equation} 
where $\tau_q= m_H^2/(4 m_q^2)$ with $m_q$ denoting any heavy-quark
mass.  The function $F_0(\tau_q)$ is defined
as~\cite{Spira:1995rr,Spira:1997dg,Harlander:2005rq}
\begin{equation}
\label{eq:f0tau}
F_0(\tau_q) = \frac{3}{2\tau_q^2}(\tau_q+(\tau_q -1)f(\tau_q)),
\end{equation}
where 
\begin{displaymath}
f(\tau_q) = \left\{ \begin{array}{ll}
\arcsin^2(\sqrt{\tau_q}), \,\,& \tau_q\leq 1\\
-\frac{1}{4}\left[\ln\frac{\sqrt{\tau_q}+\sqrt{\tau_q-1}}{\sqrt{\tau_q}-\sqrt{\tau_q-1}}-i\pi\right]^2, \,\,& \tau_q > 1.
\end{array} \right.
\end{displaymath}
The one loop virtual corrections were computed in
refs.~\cite{Spira:1995rr,Spira:1997dg,Harlander:2005rq}. In our
resummation formula~\eqref{eq:SigmaNNLL-result} they have the form
\begin{equation}
\label{eq:H1}
 \mathcal{H}^{(1)} = H^{(1)}- \left( B^{(1)}+\frac{A^{(1)}}{2}\ln{\frac{m_H^2}{Q^2}}\right)\ln{\frac{m_H^2}{Q^2}}
 + 4\pi\beta_{0}\ln{\frac{\mu_{R}^2}{m_H^{2}}}\, ,
\end{equation}
where
\begin{equation}
 H^{(1)} = 2\, c^{\rm H}(m_t,m_b) + \left(2+\frac{C_A}{2}\right)\pi^2\,,
\end{equation}
where $\beta_0 = (11 C_A-2 n_f)/(12\pi)$ and $C_A = N_c$.  All the
remaining coefficients in eq.~\eqref{eq:H1} are defined in
ref.~\cite{Banfi:2012jm}, while the expression we use for $c^{\rm
  H}(m_t,m_b)$ is taken from eq.~(3.5) of
ref.~\cite{Harlander:2005rq}. For the sake of simplicity we omit the
explicit dependence of $\mathcal{H}^{(1)}$ and $H^{(1)}$ on the quark
masses.

We obtain the fixed-order predictions from the program {\tt hnnlo
  2.0}~\cite{Grazzini:2013mca}\footnote{We wish to thank M. Grazzini
  for providing us with a preliminary version of this code.}, in which
the full NLO mass dependence is implemented and ${\cal O}(\alpha_s^4)$
corrections are computed in the large-$m_t$ limit and rescaled by the
Born top-quark-mass correction factor $|F_0(\tau_t)|^2$ where
$F_0(\tau_t)$ is defined in eq.~\eqref{eq:f0tau}.

\subsection{Treatment of non-factorizing terms }
\label{sec:non-factorizing}

In this section we want to study the numerical impact of
mass-corrections motivating the validity of our resummation formula
eq.~\eqref{eq:SigmaNNLL-result} and matching schemes
eqs.~\eqref{eq:scheme-a}, \eqref{eq:scheme-b} and
\eqref{eq:scheme-c}. We also study the impact of non-factorizing terms
and give a procedure to estimate the uncertainty associated to them.

Since we can compute mass-effects exactly only to first order in
perturbation theory, we study the impact of mass-effects on the
so-called remainder function.  The remainder function is obtained from
the fixed-order result, once all divergent logarithms are subtracted
at the given order in perturbation theory. In our case, we treat all
logarithms of the form $\ln(p_t/m_H)$, that can be resummed using
standard techniques, as divergent, and hence we subtract them from the
fixed order to obtain the remainder function.\footnote{More precisely,
  we subtract $\tilde L$ terms defined in eq.~\eqref{eq:Ltilde},
  rather then $L$.} For bottom induced production, defined as both the
top-bottom interference and the bottom squared contribution, the
remainder function will still contain logarithms of the form
$\ln(p_t/m_b)$ for $p_t > m_b$, which vanish when $p_t \to 0$.
Nevertheless, even if these logarithms vanish for $p_t\to 0$ one
should keep in mind that additional large logarithms of the ratio
$m_b/m_H$ are present in the virtual corrections Eq.~\eqref{eq:H1}
(see e.g. Eq. (44) of~\cite{Spira:1995rr}). At finite $\ptjv$
logarithms $\ln^2(m_b/m_H)$ are also present in the remainder
function. Since we do not know the higher-order structure of such
terms, in the following we will propose various ways to estimate their
impact.

\begin{figure}[htp] \centering
\includegraphics[width=0.7\columnwidth]{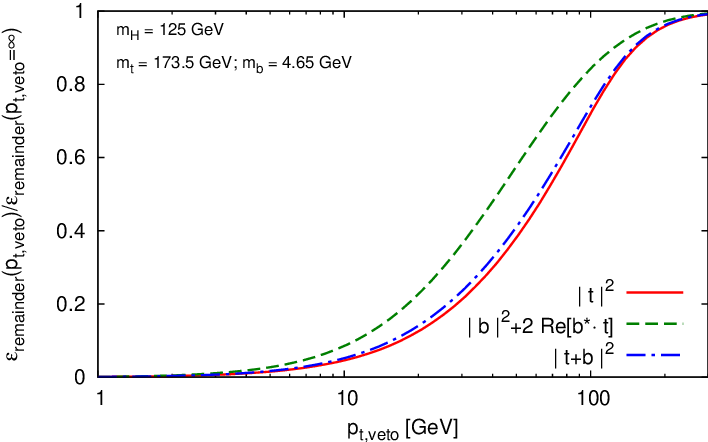}
\caption{Regular terms in the jet veto efficiency including top-quark
  effects (solid red), bottom-quark and interference effects (dashed
  green) and full top- and bottom-quark effects (dashed blue). Each
  curve is normalized to its asymptotic value at large $\ptjv$.}
  \label{fig:remainder}
\end{figure}
\begin{figure}[htp]
  \centering
  \includegraphics[width=0.48\columnwidth]{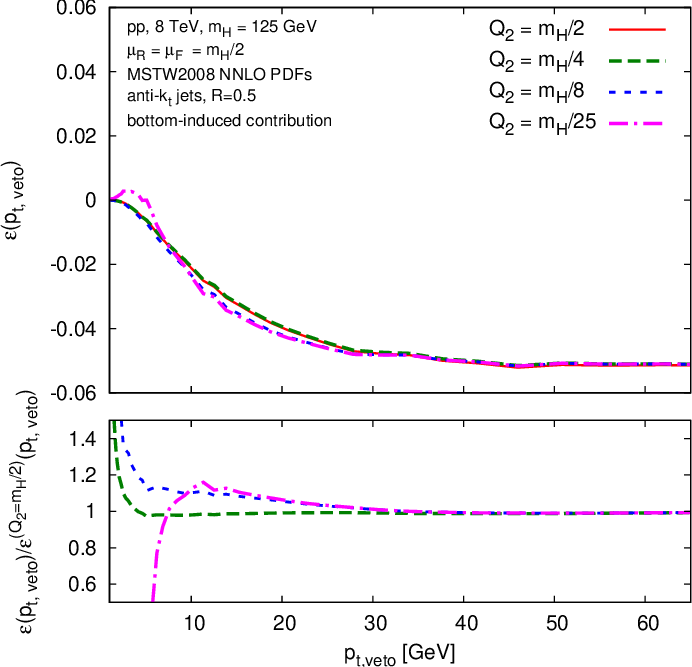}
  \includegraphics[width=0.48\columnwidth]{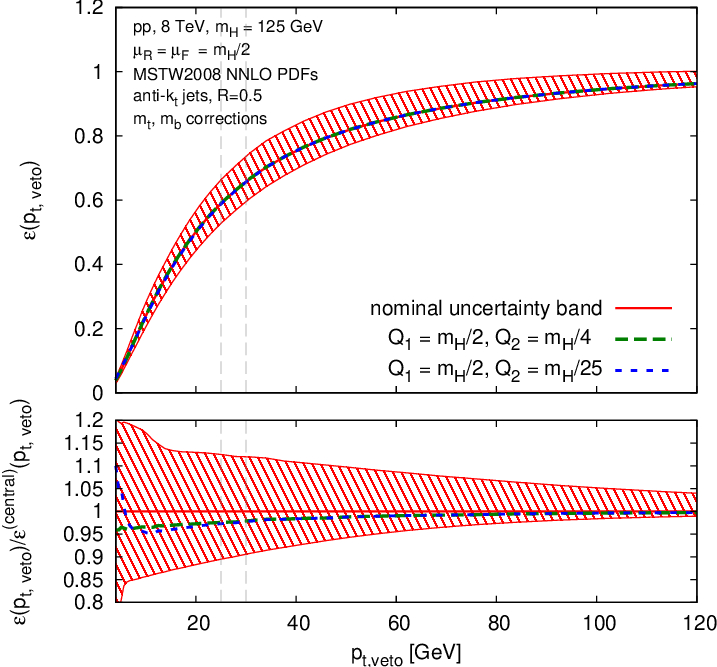}
  \caption{The effect of varying $Q_2$, while keeping all our scales
    and settings to their default values, on the bottom-induced
    contribution (left) and on the total resummed efficiency (right).}
  \label{fig:Q2}
\end{figure}

In Fig.~\ref{fig:remainder} we show the remainder contributions to the
jet-veto efficiency including top-quark effects only (solid, red),
bottom-induced (dashed, green) and full top- and bottom-quark effects
(dot-dashed, blue).
Each curve is normalized to its asymptotic value at large $\ptjv$,
this is because we are interested in studying the departure from the
infrared ($\ptjv \to 0$) limit.
From the plot it is evident that the bottom induced contribution
(dashed, green) rises faster than the pure top contribution (solid,
red). It is however also clear that once all effects are taken into
account (dot-dashed, blue) the difference with respect to the top only
contribution is modest. This is expected, since the top contribution
is dominant.  The faster rise of the bottom induced contribution can
be interpreted as an indication that the resummation starts being
unreliable at a lower $\ptjv$ value, meaning that a lower resummation
scale should be used for the bottom-induced contribution. Similarly to
what has been done in ref.~\cite{Grazzini:2013mca} we can then
introduce two different resummation scales: $Q_1$ for the top
contribution and $Q_2$ for the bottom-induced one. We show then in
Fig.~\ref{fig:Q2} the impact of varying $Q_2$ in our default matched
formula, eq.~\eqref{eq:scheme-a} with default settings as defined in
ref.~\cite{Banfi:2012jm} and recalled at the beginning of
Sec.~\ref{sec:pheno}, from values as low as the bottom mass (we choose
$m_H/25$) to values as high as half the Higgs mass.

Fig.~\ref{fig:Q2} (left) shows the effect of $Q_2$ variation on the
pure bottom-induced contribution, while keeping all other scales and
settings to their default values. In practice, we obtain the
bottom-induced contribution by subtracting the top-only correction
from the full result which includes both top and bottom effects.  We
take values of $Q_2$ equal to $m_H/25$, $m_H/8$, $m_H/4$, and $m_H/2$.
We see that, as expected, two curves obtained respectively with $Q_2 =
Q_a$ and $Q_2 = Q_b$ are in good agreement for $\ptjv > \max(Q_a,
Q_b)$. Furthermore, all curves agree above $\ptjv = 30-40$ GeV,
indicating that the resummation of bottom-induced corrections becomes
negligible in this region.  Fig.~\ref{fig:Q2} (right) shows the full
resummed efficiency including both top and bottom-induced
corrections. We see that the curves obtained with $Q_2 = m_H/25$ or
$Q_2 = m_H/4$ (while keeping again all our settings to their default
values) lie within our nominal uncertainty band, as defined at the
beginning of Sec.~\ref{sec:pheno} and shown as a red band in
Fig.~\ref{fig:Q2} (right). In particular, we notice a decrease of the
order of $2-3\%$ in the central value at $\ptjv = 25$-$30$ GeV when
taking a lower $Q_2$ value. Furthermore, we see that all predictions
lie on top of each other (i.e. differences are at the sub-permille
level) for $\ptjv$ of the order of the Higgs mass. This is expected
since our matching formulae guarantee that at high values of $\ptjv$
our predictions reproduce the fixed order result.

Another possible way to estimate the impact on the final result of
missing higher orders is to consider also a matching scheme in which
the whole one-loop remainder is exponentiated.
Matching schemes where these terms are naturally exponentiated belong
to the so called logR matching-scheme
category~\cite{Catani:1992ua}. We define a hybrid matching scheme,
where the one-loop remainder is exponentiated whilst the two-loop
contribution is treated as done in scheme (a). We refer to this scheme
as ``logR-(a)'', it is defined more precisely in
Appendix~\ref{sec:matchingschemes}. We then use the difference between
our default matching scheme ``a'' and the ``logR-(a)'' matching
scheme, to assess the uncertainty due to bottom-induced mass effects.
 \begin{figure}[htp]
  \centering \includegraphics[width=0.48\columnwidth]{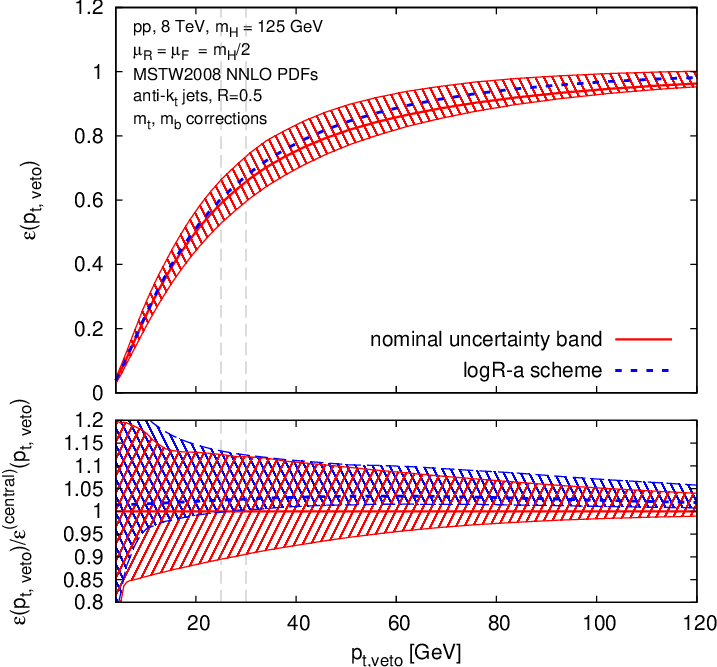}
 \includegraphics[width=0.48\columnwidth]{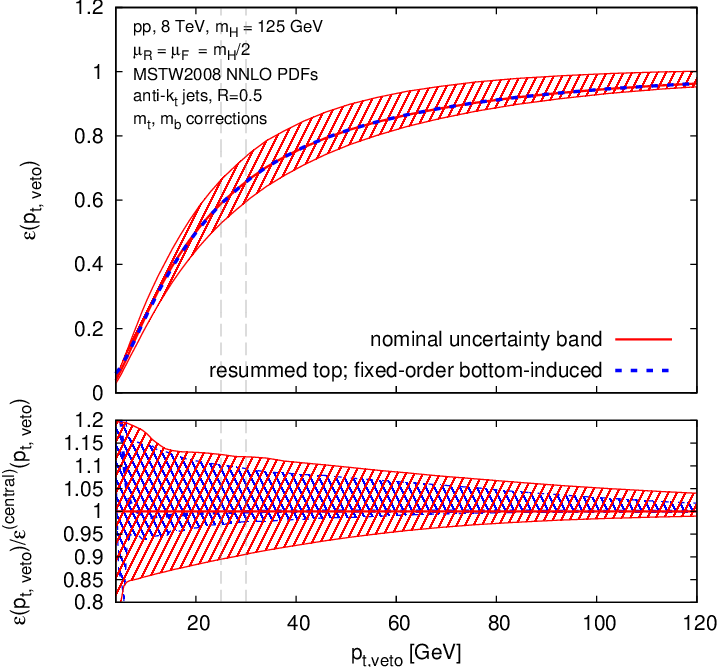}
  \caption{The left plot shows a comparison between our nominal
    uncertainty band (red) and the one obtained using the logR-(a)
    matching scheme (blue). See text for more details. The right plot
    shows the same nominal uncertainty band (red) compared to a
    prediction in which only the top-quark contributions are resummed,
    while keeping the whole bottom-induced contribution at fixed-order
    (blue). The uncertainty band in the latter case is obtained by
    varying renormalization, factorization and resummation scales.}
  \label{fig:loga-scheme}
\end{figure}

Figure~\ref{fig:loga-scheme} (left) shows the difference between our
nominal uncertainty band (defined in Section~\ref{sec:pheno}) and an
uncertainty band associated to logR-(a) scheme. The latter is obtained
by performing the same scale variations as for the nominal band, while
keeping the matching scheme fixed. We see that at $\ptjv=25$-$30$\,GeV
the central value is increased by about $2-3\%$ but it lies well
within the nominal uncertainty band.  We also note that the logR-(a)
band is smaller than the nominal band. This is not surprising since
the latter band additionally accounts for matching scheme
uncertainties. In particular, the lower edge of the nominal band is
driven by matching scheme (b).  Consistently with what we have done in
the large-$m_t$ case~\cite{Banfi:2012yh,Banfi:2012jm}, we could add
the central value of the logR-(a) scheme band to our nominal
band. However, in practice this does not make any difference since it
is fully included in the original nominal band.

The right plot of Figure~\ref{fig:loga-scheme} shows a comparison
between the nominal uncertainty band and a conservative prediction in
which only the top-quark contributions are resummed to NNLL, while the
bottom-induced contributions are simply added at fixed-order as
obtained with the generator {\tt hnnlo} 2.0. The blue uncertainty band
in the latter case is obtained as the envelope of the variation of
renormalization and factorization scales in both the fixed-order and
the resummed terms and the resummation scale in the resummed
contribution. Also in this case we do not observe sizable differences
between the two central values and the resulting uncertainty band is
fully compatible with our nominal prediction.

In a similar way, the effect of ``factorization breaking'' above $m_b$ can
be studied by introducing a modified (a) matching scheme, mod-(a), in
which the one-loop remainder is treated additively, rather than being
multiplied by a Sudakov factor. We find that the numerical difference
from scheme (a) is negligible.  A detailed discussion is reported in
Appendix~\ref{sec:modascheme}.

We therefore conclude that the method we have used to assess the theory
uncertainties in the large-$m_t$ case is robust and does not need to
be modified when mass-effects are taken into account.

\section{Phenomenology}
\label{sec:pheno}
In the present section we study the phenomenological impact of
heavy-quark mass corrections on the jet-veto efficiency and cross
section.  We consider $8$ TeV LHC collisions and the production of a
Higgs boson of mass $m_H = 125$ GeV and set the top and bottom masses
to $m_t=173.5$ GeV and $m_b = 4.65$ GeV, respectively.
Throughout our analysis we use the PDF set
MSTW2008NNLO~\cite{Martin:2009iq} and the anti-$k_t$ jet
algorithm~\cite{Cacciari:2008gp} with $R=0.5$, as implemented in
FastJet~\cite{FastJet} to reconstruct jets.

As detailed in the previous section, we match the NNLL resummed cross
section to the fixed-order NNLO prediction using the three matching
schemes~\eqref{eq:scheme-a},~\eqref{eq:scheme-b},
and~\eqref{eq:scheme-c}. We remark that the matching schemes used here
differ in one important aspect from the ones used in the large-$m_t$
case in ref.~\cite{Banfi:2012jm}. In fact, the expansion at relative
order $\cO{\alpha_s^2}$ appearing in the matching formulae is obtained
from the large-$m_t$ resummation formula~\cite{Banfi:2012jm}, rescaled
by the Born top-quark-mass correction factor $|F_0(\tau_t)|^2$ where
$F_0(\tau_t)$ is defined in eq.~\eqref{eq:f0tau}. This ensures that
the matching procedure does not spoil the NNLL accuracy of our
resummation in the small $\ptjv$ limit. However, this also implies
that the difference between matched and fixed-order results remains of
relative order $\cO{\alpha_s^2}$, and not $\cO{\alpha_s^3}$ as one
would expect from a NNLO matching.
The uncertainties in the jet-veto efficiency
and in the zero-jet cross section are assessed as previously discussed
in refs.~\cite{Banfi:2012yh,Banfi:2012jm} and as recalled here. The
default central prediction is obtained by setting renormalization
$\mu_R$, factorization $\mu_F$ and resummation scale $Q$ all equal to
half the Higgs mass, $\mu_F=\mu_R=Q=m_H/2$, and matching scheme
(a),~eq.~\eqref{eq:scheme-a}.
Uncertainties are obtained by varying both $\mu_F$ and $\mu_R$ by a
factor of two in either direction, keeping $1/2\,\leq
\,\mu_R/\mu_F\,\leq \,2$. Moreover, using central values for $\mu_R$
and $\mu_F$, we vary the resummation scale $Q$ by a factor of two in
scheme (a) and compute the central values of matching schemes (b),
eq.~\eqref{eq:scheme-b} and (c), eq.~\eqref{eq:scheme-c}.  The final
uncertainty for the matched prediction is obtained as the envelope of
all these variations.

\begin{figure}[htp]
  \centering
  \includegraphics[width=0.48\columnwidth]{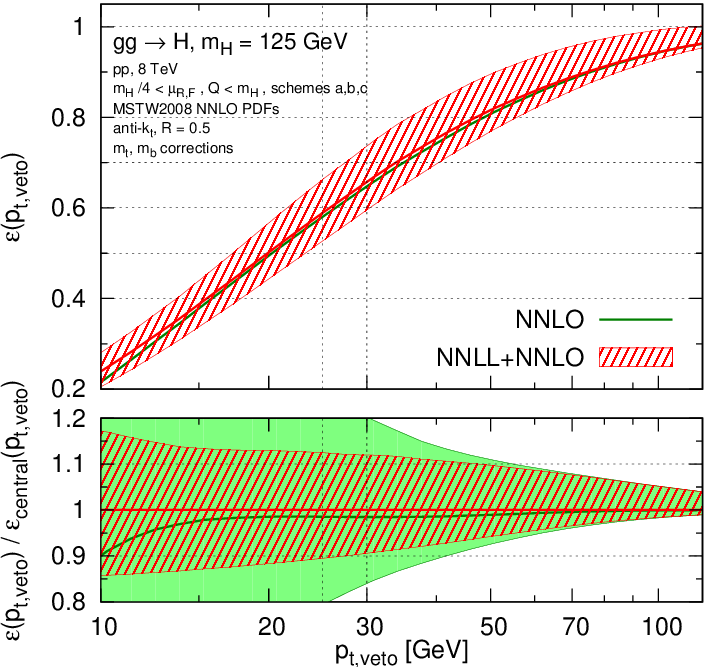}\hfill
  \includegraphics[width=0.48\columnwidth]{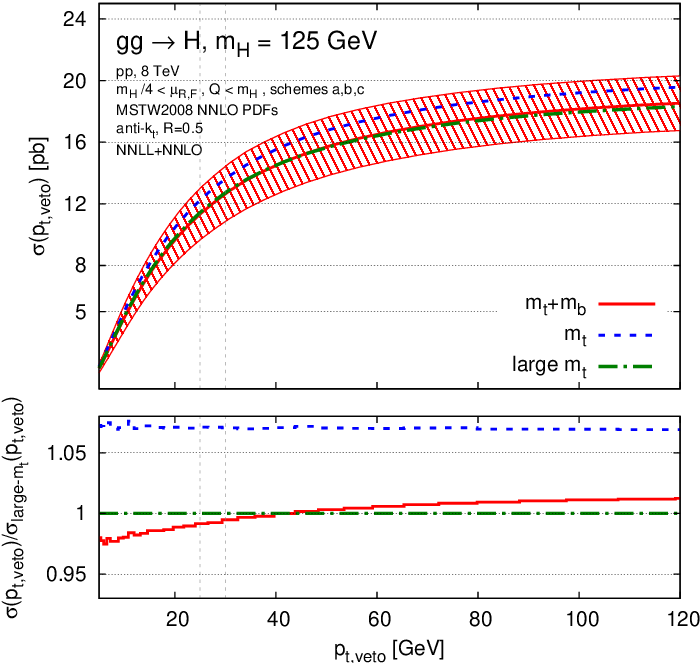}
  \caption{Left: comparison between the matched efficiency at
    NNLL+NNLO accuracy and the NNLO fixed-order result, both results
    include mass effects. Right: zero-jet cross section computed in the
    large-$m_t$ limit (green dot-dashed curve), including full $m_t$ dependence
    (blue dashed curve), and including full $m_t$ and $m_b$ dependence (red
    solid band).}
  \label{fig:plots-results}
\end{figure}

Numerical results for the jet-veto efficiency are shown in the left
plot of Fig.~\ref{fig:plots-results}. In the region of interest for
LHC analyses, i.e. $\ptjv$ in the range 25$-$30 GeV, the matched
prediction for the efficiency is 2-3\% higher than the fixed-order
one. The corresponding theoretical uncertainty is reduced by roughly a
factor of two when the resummation is included.
The right plot of Fig.~\ref{fig:plots-results} shows the comparison
between the NNLL+NNLO exclusive zero-jet cross section in the
large-$m_t$ limit and the one including the full top- and bottom-quark
mass dependence. We see that the integrated cross section is about 7\%
larger if the full top-quark mass dependence is included (quite
independently of the value of $\ptjv$), whereas it is 1-2\% lower for
$\ptjv=25$-$30$ GeV, when also the bottom is taken into account.

\begin{figure}[htp]
  \centering
  \includegraphics[width=0.70\columnwidth]{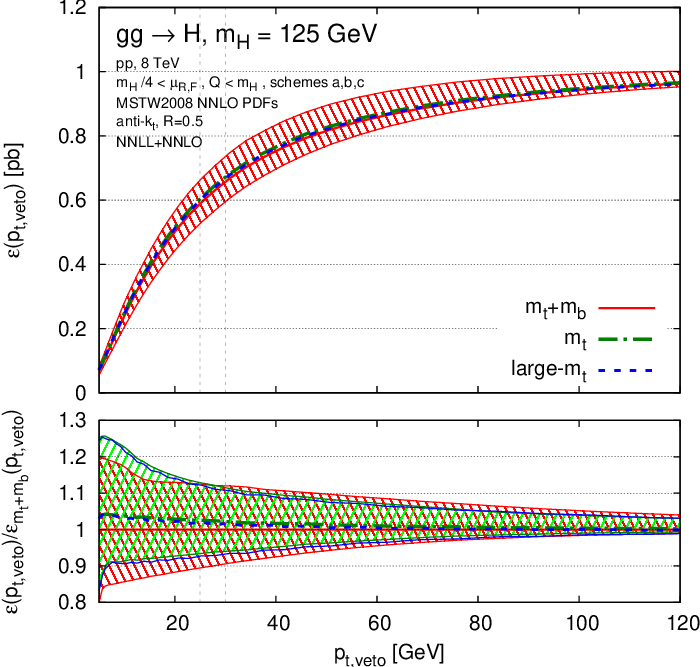}\hfill
  \caption{Comparison between the matched efficiency at NNLL+NNLO
    accuracy in the large-$m_t$ limit (dashed blue), including $m_t$ only
    (dot-dashed green) and both $m_t$, $m_b$ effects (solid red). The lower panel shows
    the ratio to the central value of the $m_t+m_b$ band.}
  \label{fig:effplot}
\end{figure}

In Fig.~\ref{fig:effplot} we show the jet-veto efficiency computed at
NNLL+NNLO with and without exact mass dependence. We observe an
increase of $2$-$3\%$ in the uncertainty at $\ptjv=25$-$30$\,GeV when
the bottom-induced contributions are included, compared to both the
large and the exact $m_t$ results. This is due to the larger remainder
function which pushes the schemes (b) and (c) results to lower and
higher values, respectively. As a consequence, the uncertainty band in
the presence of full mass corrections is totally driven by the central
values obtained with schemes (b) and (c) which constitute the lower
and upper edges of the band, respectively. This is in contrast with
what happens both in the large-$m_t$ limit and if only $m_t$
corrections are included where the upper edge of the band is driven by
the resummation scale variation.

Below we provide tables with numerical results for cross-sections and
efficiencies for the values of veto scales and jet radii, as used in
current LHC analyses. All uncertainties have been symmetrized with
respect to the central value.


\begin{table}
\begin{center}
  \center{\bf{Exact $\bf m_t$ and $\bf m_b$ corrections}}\\ 
  \vspace{0.8em}
  \begin{tabular}{c|c|c|c|c}
        R & $\,\ptjv$\,[GeV] & $\sigma_{\rm tot}^{{8\TeV}}$\,[pb] & 
    $\epsilon^{(8 \TeV)}$ & \,$\sigma^{(8\TeV)}_\text{0-jet}$\,[pb]
    \\[0.2em]\hline
    \phantom{x} & & & &
    \\[-1em] 
    $0.4$\;& $25$ & \,$19.24\,\pm\,1.78$\, & 
    $0.602\,\pm\,0.070$ & $11.59\,\pm\,1.72$ 
    \\[0.4em] 
    $0.5$\;& $30$ & \,$19.24\,\pm\,1.78$\, &
    \,$0.657\,\pm\,0.070$\, & $12.64\,\pm\,1.79$ 
  \end{tabular}
\end{center}
  \caption{Total inclusive cross-section, jet-veto efficiency and
    zero-jet cross-section for Higgs production at the 8 TeV LHC for two
    different values of the jet radius $R$ and $\ptjv$. Results
    include exact top and bottom mass dependence. The quoted total
    cross section and the corresponding errors have been computed with
    the {\tt hnnlo} 2.0 code~\cite{Catani:2007vq}. }
\label{tab:mtmb}
\end{table}

\begin{table}
\begin{center}
  \center{\bf{Exact $\bf m_t$ corrections only}}\\ 
  \vspace{0.8em}
  \begin{tabular}{c|c|c|c|c}
        R & $\,\ptjv$\,[GeV] & $\sigma_{\rm tot}^{{8\TeV}}$\,[pb] & 
    $\epsilon^{(8 \TeV)}$ & \,$\sigma^{(8\TeV)}_\text{0-jet}$\,[pb]
    \\[0.2em]\hline
    \phantom{x} & & & &
    \\[-1em] 
    $0.4$\;& $25$ & \,$20.16\,\pm\,1.87$\, & 
    $0.617\,\pm\,0.063$ & $12.44\,\pm\,1.72$ 
    \\[0.4em] 
    $0.5$\;& $30$ & \,$20.16\,\pm\,1.87$\, &
    \,$0.672\,\pm\,0.057$\, & $13.54\,\pm\,1.71$ 
  \end{tabular}
\end{center}
  \caption{As table~\ref{tab:mtmb} but including only exact top mass
    dependence.}
\label{tab:mt}
\end{table}

\begin{table}
\begin{center}
  \center{\bf{Large-$\bf m_t$ approximation}}\\ 
  \vspace{0.8em}
  \begin{tabular}{c|c|c|c|c}
        R & $\,\ptjv$\,[GeV] & $\sigma_{\rm tot}^{{8\TeV}}$\,[pb] & 
    $\epsilon^{(8 \TeV)}$ & \,$\sigma^{(8\TeV)}_\text{0-jet}$\,[pb]
    \\[0.2em]\hline
    \phantom{x} & & & &
    \\[-1em] 
    $0.4$\;& $25$ & \,$19.03\,\pm\,1.76$\, & 
    $0.613\,\pm\,0.064$ & $11.66\,\pm\,1.62$ 
    \\[0.4em] 
    $0.5$\;& $30$ & \,$19.03\,\pm\,1.76$\, &
    \,$0.667\,\pm\,0.058$\, & $12.70\,\pm\,1.61$ 
  \end{tabular}
\end{center}
  \caption{As table~\ref{tab:mtmb} but in the large $m_t$ approximation.}
\label{tab:largemt}
\end{table}

\begin{table}
\begin{center}
  \center{\bf{Large-$\bf m_t$ approximation ($\sigma_{\rm tot}^{8\TeV}$ from HXSWG)}}\\ 
  \vspace{0.8em}
  \begin{tabular}{c|c|c|c|c}
        R & $\,\ptjv$\,[GeV] & $\sigma_{\rm tot}^{{8\TeV}}$\,[pb] & 
    $\epsilon^{(8 \TeV)}$ & \,$\sigma^{(8\TeV)}_\text{0-jet}$\,[pb]
    \\[0.2em]\hline
    \phantom{x} & & & &
    \\[-1em] 
    $0.4$\;& $25$ & \,$19.27\,\pm\,1.45$\, & 
    $0.613\,\pm\,0.064$ & $11.81\,\pm\,1.51$ 
    \\[0.4em] 
    $0.5$\;& $30$ & \,$19.27\,\pm\,1.45$\, &
    \,$0.667\,\pm\,0.058$\, & $12.86\,\pm\,1.47$ 
  \end{tabular}
\end{center}
  \caption{As table~\ref{tab:mtmb} but in the large $m_t$
    approximation. Unlike in table~\ref{tab:largemt} the total
    cross-section is taken from the HXSWG~\cite{Heinemeyer:2013tqa}
    and includes finite-width, electro-weak and threshold resummation
    effects.}
\label{tab:largemthxswg}
\end{table}
\vspace{0.7cm}

We first examine only the first three tables in which the total
cross-section is computed at NNLO QCD. We notice that top-mass effects
increase the central values by about 6-7\%. Bottom mass effects bring
the central values down again, leading to a modest decrease of about
half a percent in the zero-jet cross-section with respect to the
large-$m_t$ result. As already noticed earlier, we see that the
uncertainties in the cross-section remain unchanged when including
only top mass effects, while they increase by about 2\% (and amount to
about 14\%) when also bottom-induced corrections are taken into
account.
By comparing with Fig.~2 of ref.~\cite{Banfi:2012jm} we note that when
mass corrections are taken into account, the uncertainty in the
resummed prediction increases by about 2\%, while the fixed order
uncertainty increases from about 15-20\% to about 20-25\%.

In tab.~\ref{tab:largemthxswg} we report the jet-veto efficiency and
cross-section using the improved total cross-section recommended by
the Higgs Cross Section Working Group
(HXSWG)~\cite{Heinemeyer:2013tqa} instead of the pure NNLO
value. Improvements include the treatment of the Higgs width, NNLL
threshold effects and NLO electro-weak corrections. To figure out only
the size of mass effects one has to compare results including mass
corrections to tab.~\ref{tab:largemt}, rather than
tab.~\ref{tab:largemthxswg}. The improved predictions for the total
cross section included in tab.~\ref{tab:largemthxswg} are not
available when finite-mass effects are included. The difference
between tables~\ref{tab:largemt} and~\ref{tab:largemthxswg} can then
help estimate the effect of these higher-order corrections on the
results of tables~\ref{tab:mtmb} and \ref{tab:mt}.

\begin{figure}[htp]
  \centering
  \includegraphics[width=0.48\columnwidth]{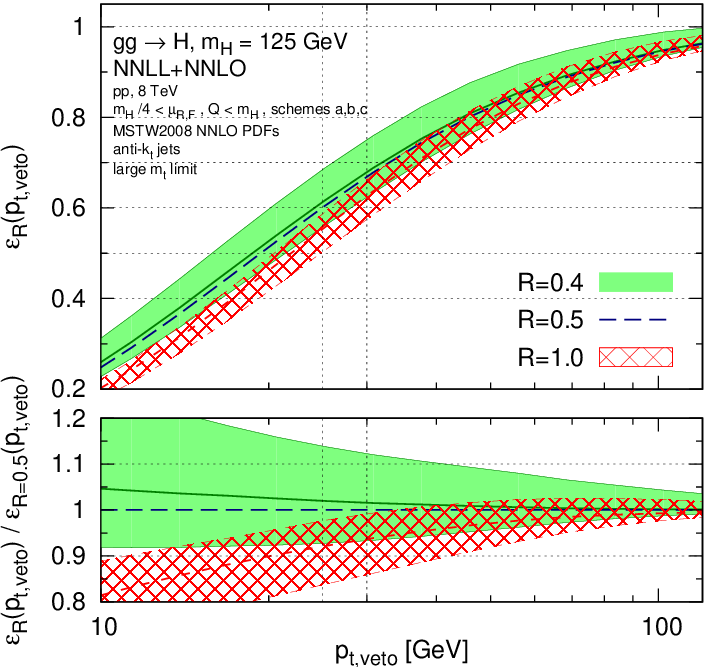}
  \includegraphics[width=0.48\columnwidth]{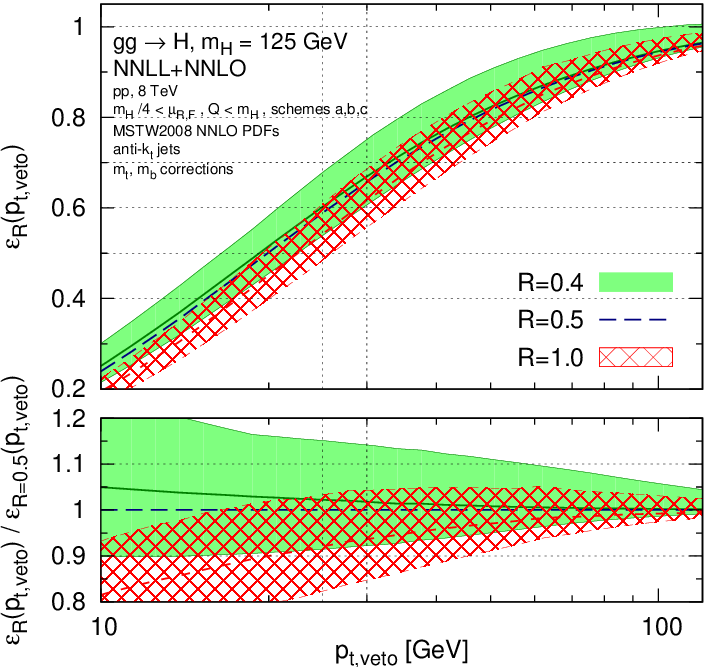}
  \caption{Comparison between the matched efficiencies at NNLL+NNLO
    accuracy for different values of the jet radius in the large-$m_t$
    limit (left) and with $m_t$, $m_b$ effects (right). The lower
    panel shows the ratio to the central value with $R=0.5$.}
  \label{fig:threeR}
\end{figure}

All results presented so far used $R=0.5$ as a jet-radius to define
jets.  As noticed first in ref.~\cite{Banfi:2012jm}, when the Higgs is
produced in the large-$m_t$ limit the uncertainty band is reduced when
using a larger radius.  This is because in the latter case the upper
edge of the uncertainty band is determined by resummation scale
dependence of the NNLL corrections which are smaller at larger values
of $R$. As we pointed out previously, when we include the exact mass
dependence the uncertainty band is fully determined by the spread
between matching schemes (b) and (c). This spread is not expected to
decrease when larger values of $R$-values are considered, this is why
we do not observe any reduction in the uncertainty when considering a
larger radius (see Fig.~\ref{fig:threeR}).

\begin{figure}[htp]
  \centering
  \includegraphics[width=0.48\columnwidth]{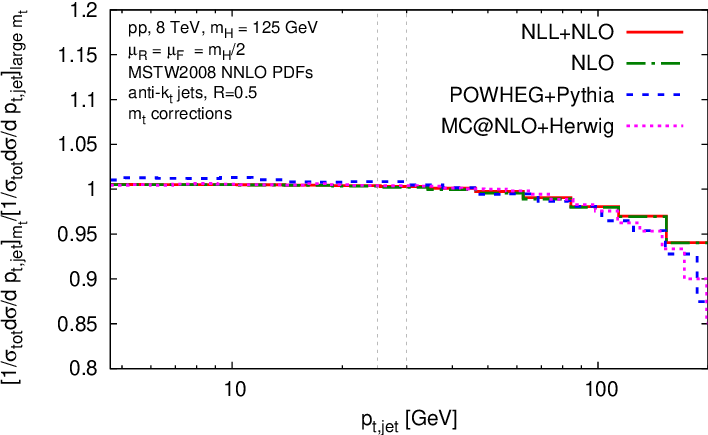}\hspace{1em}
  \includegraphics[width=0.48\columnwidth]{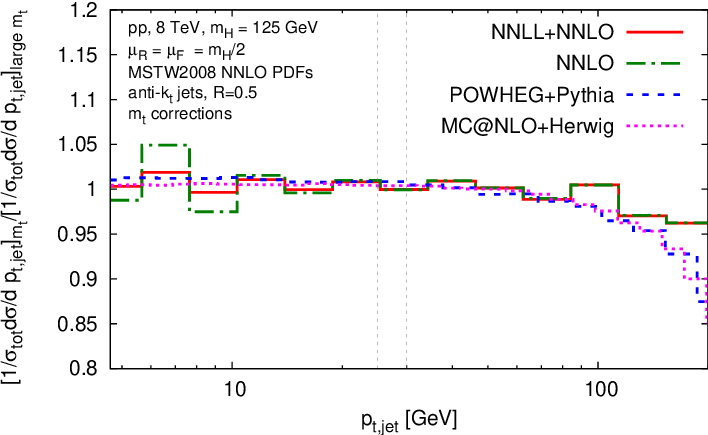}
  \caption{Ratios of the leading-jet $p_t$ distribution (normalized to
    the total cross section) including full dependence on the top
    mass, to the same distribution in the large-$m_t$
    approximation (also normalized). In the plots labelled NNLL+NNLO and NNLO, mass
    corrections are included only at NLO, as described in the text.}
  \label{fig:comparison-to-MC-mt}
\end{figure}

\begin{figure}[htp]
  \centering
  \includegraphics[width=0.48\columnwidth]{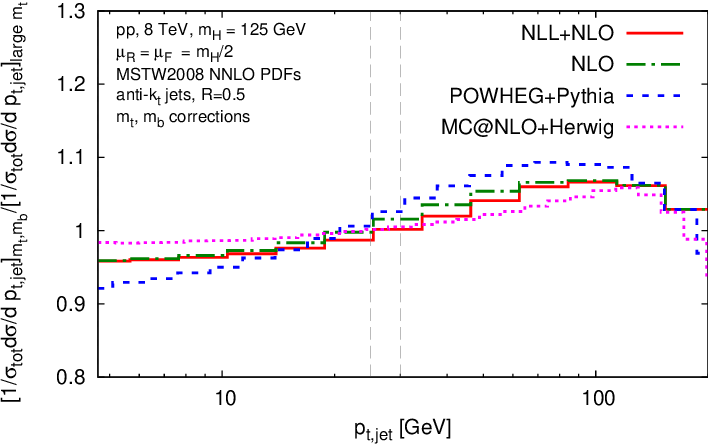}\hspace{1em}
  \includegraphics[width=0.48\columnwidth]{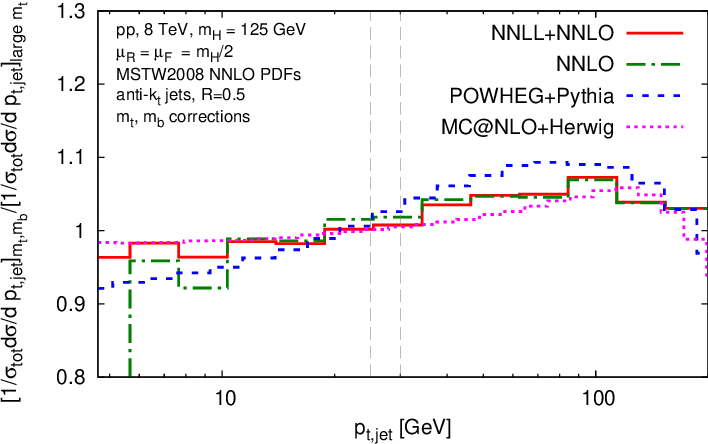}
\caption{As in Fig.~\ref{fig:comparison-to-MC-mt} including dependence
  on top and bottom masses.}
  \label{fig:comparison-to-MC-mtmb}
\end{figure}

Since mass effects are also implemented in NLO Monte Carlo generators,
{\tt POWHEG}~\cite{Nason:2004rx,Alioli:2010xd,Bagnaschi:2011tu} and
{\tt MC@NLO}~\cite{Frixione:2002ik,mcatnlo}, it is instructive to
carry out a comparison to their predictions. The comparison is
particularly interesting since the two event generators have a
different treatment of heavy-quark mass corrections. In particular, in
{\tt POWHEG} mass corrections are included in the Sudakov form factor,
while in {\tt MC@NLO} they are treated as finite remainders.
The right plots of Figs.~\ref{fig:comparison-to-MC-mt} and
~\ref{fig:comparison-to-MC-mtmb} show four different predictions for
the ratio of the leading-jet $p_t$ distribution, normalized to the
corresponding total cross section, to the same distribution in the
large-$m_t$ approximation, as obtained from {\tt JetVHeto} at
NNLL+NNLO (red, solid), at NNLO (green, dot-dashed), {\tt POWHEG+Pythia}
(blue, dashed) and {\tt MC@NLO+Herwig} (red, dashed).  All Monte
Carlos are run at parton level only, with no multi-parton interactions
or hadronization corrections. For completeness, the comparison to
NLL+NLO and NLO is reported in the left plots of
Figs.~\ref{fig:comparison-to-MC-mt}
and~\ref{fig:comparison-to-MC-mtmb}.

We see that the three predictions for the ratio agree well if only the
top-quark is included (Fig.~\ref{fig:comparison-to-MC-mt}). At high
$\ptjv$ {\tt JetVHeto} differs from the NLO Monte Carlo
predictions in the right plots. This is not surprising since {\tt JetVHeto} is NLO
(rather then LO) accurate in the jet-veto spectrum.  On the contrary,
when bottom-quark effects are included
(Fig.~\ref{fig:comparison-to-MC-mtmb}), predictions differ over the
whole spectrum.

In general we find that in this case the prediction from {\tt
  JetVHeto} lies somewhat in between that of {\tt POWHEG+Pythia} and
{\tt MC@NLO+Herwig}, but tends to be closer to the latter.
In particular, at usual veto scales, $25\,{\rm
  GeV}\,\leq\,\ptjv\,\leq\,30\,{\rm GeV}$, we found better agreement
with {\tt MC@NLO}.  Compared to {\tt JetVHeto}, {\tt POWHEG} seems to
enhance the size of $m_b$ effects, while {\tt MC@NLO} seems to
diminish them.

\begin{figure}[htp]
  \centering
  \includegraphics[width=0.70\columnwidth]{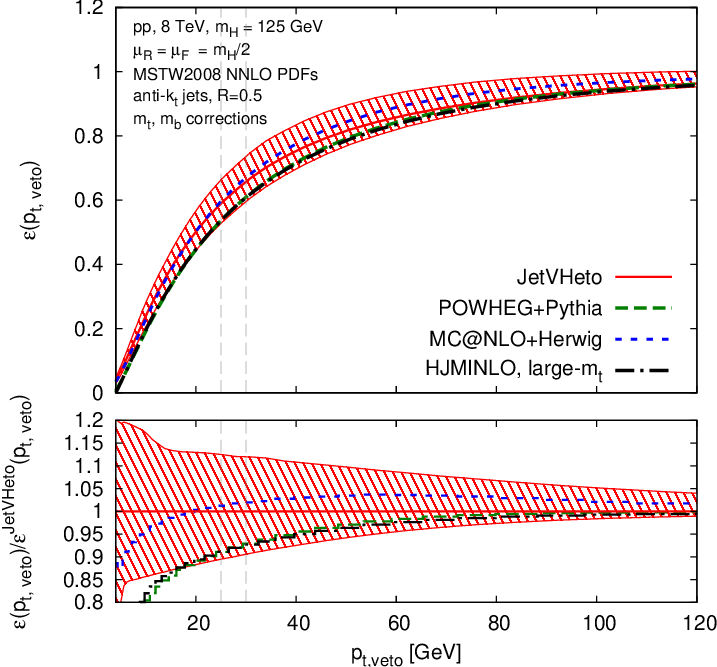}
  \caption{Comparison between different event generators for the jet-veto efficiency.}
  \label{fig:distributions-efficiency}
\end{figure}

Finally, it is interesting to verify whether Monte Carlo predictions
lie in the uncertainty band of {\tt JetVHeto} for the efficiency.
Fig.~\ref{fig:distributions-efficiency} shows the predictions for the
jet-veto efficiency obtained with {\tt JetVHeto}, with its uncertainty
band, {\tt POWHEG}, {\tt MC@NLO} and {\tt
  HJ-MiNLO}~\cite{Hamilton:2012np,Hamilton:2012rf} (the latter uses
the large $m_t$ approximation). We find that for $\ptjv > 20$ GeV all
predictions lie in the uncertainty band of {\tt JetVHeto}.
In fact, {\tt POWHEG+Pythia} tends to the central value of the {\tt
  JetVHeto} predictions at high $\ptjv$, while {\tt MC@NLO} is closer
to {\tt JetVHeto} at lower $\ptjv$.

\section{Conclusions}
\label{sec:conclu}

In the present work we studied the size of finite-mass effects in the
resummed jet-veto efficiency and zero-jet cross section for Higgs-boson
production. The inclusion of these corrections is not trivial since
the mass of virtual quarks introduce additional scales in the problem,
besides the Higgs mass and the jet-veto scale.  In particular, when
the bottom quark is included, new non-factorizing logarithms of the
type $\ln(\ptjv/m_b)$ appear if the emitted final state partons
resolve the quark loop, i.e. in the kinematical range $m_b < \ptjv$.
 
Since such new logarithms vanish for $\ptjv < m_b$, we argue that it
is reasonable to treat them as any regular remainder.
We have validated our resummation and matching procedure by varying
the resummation scale related to the bottom-induced terms, and by
exponentiating the one-loop remainder using a logR-type matching
scheme. Still, a two-loop result of mass-effects would be useful to
get an insight into the structure of non-factorizing terms to higher
orders.

As already observed in ref.~\cite{Mantler:2012bj} for the Higgs
transverse momentum distribution, we also find that bottom-mass
effects distort the jet-veto distribution. However, we also find that
bottom-quark effects are as large as finite top corrections and
opposite in sign, so that this destructive interference leads to very
small effects for jet-veto values currently used at the LHC. Compared
to the large $m_t$ limit the jet-veto efficiency (zero-jet
cross-section) decreases by about 1.5\% (0.5\%) for $\ptjv = 25$-$30$
GeV, when both top and bottom effects are taken into account. The
uncertainty however increases by about 2\% with respect to the
large-$m_t$ limit.
Unlike what observed in ref.~\cite{Grazzini:2013mca}, we find that our
predictions are stable when varying the resummation scale $Q_2$ of the
bottom-induced contribution, also at high values of
$Q_2$. Furthermore, we find modest corrections even at very small
transverse momentum values (we stress however that we are looking at
different observables).

Given the very small effects that we find here, it is natural to
wonder why mass-effects in {\tt POWHEG} or {\tt HRes}, as implemented
in ref.~\cite{Grazzini:2013mca} and ~\cite{vicinitalk}, give rise to
much larger corrections at low $p_t$. In order to investigate this
issue, we have performed a NNLL+NNLO calculation for $p_{tH}$ as well,
including mass effects as described in the present paper. Our results
are given in appendix~\ref{app:ptjvspth}. We find that, if we match
resummed results to NLO distributions, in the intermediate $p_t$
region, between 20 and 50 GeV, the matched distributions differ
substantially from the corresponding fixed-order predictions when
choosing a standard resummation scale of the order of teh Higgs
mass. This is true both in the large-$m_t$ limit and including
finite-mass effects (top and bottom contributions). On the other hand,
when the matching is performed at NNLO, we find a much better
agreement between matched and fixed-order distributions, more details
are given in Appendix~\ref{app:ptjvspth}. We note that
ref.~\cite{Grazzini:2013mca} considers a matching to NLO only for the
bottom-induced contribution. Insofar, our results are not in
disagreement with that reference.


The effects that we find here are very modest, because of both the
Yukawa suppression for bottom quarks and of the destructive
interference. When the leading jet has a transverse momentum that is
much larger than the top mass, as in boosted studies, logarithms of
the form $\ln(p_t/m_t)$ can be potentially large and might need to be
resummed to all orders. We stress that these mass effects are not
treated properly by Monte Carlo generators. Moreover, the
non-factorizing logarithms are not resummed by any parton shower.
When integrated over the whole $\ptjv$ spectrum, such logarithms give
rise to terms of the form $\ln(m_H/m)$ in the total cross section
which can be sizable in heavy-Higgs searches.

Our formulae have been implemented in the public code {\tt
  JetVHeto}~\cite{jetvheto}, distributed together with a number of
benchmark results.

\section*{Acknowledgments}
We particularly thank Paolo Nason and Gavin Salam for extensive
discussions. We also acknowledge useful discussions with Mrinal
Dasgupta, Keith Hamilton and Paolo Torrielli.  We thank ESI (all) and
KITP (GZ) for hospitality while part of this work was carried out. GZ
is supported by Science and Technology Facility Council. PFM is
supported by the Swiss National Science Foundation (SNF) under grant
200020-138206. AB is supported by the Science Technology and
Facilities Council (STFC) under grant number ST/J000477/1.

\appendix

\section{Real emission matrix elements}
\label{sec:app-real}

We report here the lowest order matrix elements including finite mass
effects for the production of a Higgs boson in association with a
(anti-)quark or a gluon.  These are taken directly from
ref.~\cite{Baur:1989cm}.

We denote by $p_1$ and $p_2$ the momenta of the incoming partons and
with $p_3$ that of the outgoing parton. We express the matrix elements in
terms of the usual Mandelstam invariants
\begin{equation}
  \label{eq:invariants}
  s= (p_1+p_2)^2\,,\quad t=(p_2-p_3)^2\,,\quad u=(p_1-p_3)^2\,,
\end{equation}
and of the auxiliary invariants
\begin{equation}
  \label{eq:new-invariants}
  s_1= s-m_H^2\,,\quad t_1=t-m_H^2\,,\quad u_1=u-m_H^2\,.
\end{equation}
We have four contributing subprocesses, $gg \to Hg$, $qg \to Hq$, $gq
\to Hq$, $q\bar q \to Hg$.

\subsubsection*{The subprocess $gg\to Hg$}

The helicity summed (unaveraged) amplitude squared is given by
\begin{equation}
|{\cal M}_{gg\to Hg}|^2 = \frac{N_c (N_c^2-1)}{64 \pi}\alpha_s^3
  \sum_{\lambda_1,\lambda_2,\lambda_3=\pm } \left| \sum_f M_{gg\to
    Hg}^{\lambda_1\lambda_2\lambda_3}(s,t,u)\right|^2\,, 
\end{equation}
where the sum is over different quark flavours $f=t,b$  in the loop and
$\lambda_i$ is the helicity of gluon $p_i$. 

For a quark of mass $m$ in the loop, coupled to the Higgs with a
Yukawa coupling $y=g_w m/(2 M_{\small W})$ we have:
\begin{equation}
  \label{eq:M+++}
  \begin{split}
    \frac{M_{gg\to Hg}^{+++}(s,t,u)}{y\, m\,\Delta}&=-64\left(\frac{1}{ut}+\frac{1}{t
      t_1}+\frac{1}{u
      u_1}\right)-\frac{64}{s}\left(\frac{2s+t}{u_1^2}B_1(u)+\frac{2s+u}{t_1^2}B_1(t)\right)\\
  &-\frac{16(s-4m^2)}{stu}\left[s_1
    C_1(s)+(u-s)C_1(t)+(t-s)C_1(u)\right]\\
&-128 m^2 \left(\frac{1}{t t_1} C_1(t)+\frac{1}{u
    u_1}C_1(u)\right)+\frac{64 m^2}{s}D(u,t)\\
&+\frac{8(s-4m^2)}{stu}\left[st D(s,t)+usD(u,s)-utD(u,t)\right]-\frac{32}{s^2}E(u,t)\,,
  \end{split}
\end{equation}
and
\begin{equation}
  \label{eq:M++-}
  \begin{split}
    \frac{M_{gg\to Hg}^{++-}(s,t,u)}{y\, m\,\Delta}&= \frac{64
      m_H^2}{stu}+\frac{16(m_H^2-4m^2)}{stu}
\left[s_1 C_1(s)+t_1 C_1(t)+u_1C_1(u)\right]\\
&-\frac{8(m_H^2-4m^2)}{stu}\left[st D(s,t)+usD(u,s)+utD(u,t)\right]\,,
  \end{split}
\end{equation}
where $\Delta=\sqrt{(stu)/8}$.
$B_1(s)$ is then defined as:
\begin{equation}
  \label{eq:B1}
  B_1(s) = B(s)- B(m_H^2)\,,
\end{equation}
where  $B$ is the scalar triangle integral
\begin{equation}
  \label{eq:bubble}
\begin{split}
  B(q^2)&=\int \frac{d^4 \ell}{i \pi^2} \frac{1}{[\ell^2-m^2] [(\ell+q)^2-m^2]} \\
&=-\int_0^1 dx
  \ln\left[m^2-i\epsilon-q^2x(1-x)\right]\,. 
\end{split}
\end{equation}
 $C$ is the scalar triangle integral
\begin{equation}
  \label{eq:triangle}
\begin{split}
  C(q^2)&=\int \frac{d^4 \ell}{i \pi^2} \frac{1}{[\ell^2-m^2] [(\ell+p_1)^2-m^2][(\ell+p_1+p_2)^2-m^2]} \\
&=\int_0^1\frac{dx}{q^2 x}
  \ln\left[1-i\epsilon-\frac{q^2}{m^2}x(1-x)\right]\,,\qquad
  q^2=(p_1+p_2)^2\,,
\end{split}
\end{equation}
$C_1$, $D$ and $E$ are other scalar integrals. $C_1(s)$ is defined as:
\begin{equation}
  \label{eq:C1}
  s_1 C_1(s) = s C(s)-m_H^2 C(m_H^2)\,,
\end{equation}
and an analogous definition holds for $C_1(t)$ and $C_1(u)$.

$D(s,t)$ is the box
integral
\begin{equation}
  \label{eq:box}
\begin{split}
  D(s,t) &=\int \frac{d^4 \ell}{i \pi^2} \frac{1}{[\ell^2-m^2]
    [(\ell+p_1)^2-m^2][(\ell+p_1+p_2)^2-m^2] [(\ell-p_H)^2-m^2]} \\
&=\frac{1}{st}\int_0^1 \frac{1}{x(1-x)+m^2 u/(ts)} \left\{
-\ln\left[1-i\epsilon-\frac{m_H^2}{m^2}x(1-x)\right]\right.\\ 
&\qquad \left.+\ln\left[1-i\epsilon-\frac{s}{m^2}x(1-x)\right]+\ln\left[1-i\epsilon-\frac{t}{m^2}x(1-x)\right]
\right\}\,,
\end{split}
\end{equation}
and analogous definitions hold for $D(u,s)$ and $D(t,u)$.

$E$ is the following linear combination of $C$ and $D$ integrals
\begin{equation}
  \label{eq:E}
  E(u,t)=uC(u)+tC(t)+u_1C_1(u)+t_1 C_1(t)-utD(u,t)\,.
\end{equation}

The other amplitudes are obtained from these by exchanging the
invariants, namely
\begin{equation}
  \label{eq:M-gluons}
  \begin{split}
      M_{gg\to Hg}^{-+-}(s,t,u)&=M_{gg\to Hg}^{+++}(t,s,u)\,,\\
      M_{gg\to Hg}^{-++}(s,t,u)&=M_{gg\to Hg}^{+++}(u,t,s)\,,
  \end{split}
\end{equation}
and the last four helicity amplitudes can be obtained using parity
conservation 
\begin{equation}
  \label{eq:M-gluonsP}
M_{gg\to Hg}^{\lambda_1\lambda_2\lambda_3}(s,t,u)=- M_{gg\to
  Hg}^{-\lambda_1-\lambda_2-\lambda_3}(s,t,u)\,.
\end{equation}

\subsubsection*{Amplitudes for $q \bar q \to Hg$}

The amplitude squared for the process $q\bar q \to H g$ is given by
\begin{equation}
  \label{eq:Mqq2}
 |M_{q\bar q\to Hg}|^2(s,t,u) = \frac{N_c^2-1}{2} \frac{\alpha_s^3}{\pi} \frac{t^2+u^2}{s_1^2} \frac{1}{s}
 |\mathcal{A}(s,t,u)|^2\,,
\end{equation}
where
\begin{equation}
  \label{eq:Aquark}
  \mathcal{A}(s,t,u) = \sum_f y_f\,m_f\left(2+\frac{2s}{s_1} B_1(s)+\left(4 m_f^2-t-u\right)C_1(s)\right)\,,
\end{equation}
where $m_f$ denotes the (top or bottom) mass in the loop, $y_f$ the
corresponding quark coupling to the Higgs, whereas external quarks are
treated as massless.

The matrix elements for the crossed processes $qg\to Hg$ and $g \bar q
\to H\bar q$ can be obtained from $|M_{q\bar q}|^2$ as follows
\begin{equation}
  \label{eq:Mqg-Mgqbar}
  \begin{split}
  |M_{qg\to Hq}|^2(s,t,u) &= -|M_{q\bar q \to Hg}|^2(u,t,s)\,,\\ 
  |M_{g\bar q\to H\bar q}|^2(s,t,u) &= -|M_{q\bar q \to Hg}|^2(t,s,u)\,.     
  \end{split}
\end{equation}

\section{Matching schemes}
\label{sec:matchingschemes}

In this section we report the expressions for the three matching
schemes at NNLL+NNLO defined in ref.~\cite{Banfi:2012jm} and used in this
paper.
For the purpose of this paper we define
\begin{align}
\tilde{\Sigma}_{\rm NNLL}^{(2)}(p_{\rm t, veto}) = \Sigma_{\rm NNLL}^{(2)}(p_{\rm t, veto})\bigg|_{{\rm large}-m_t}  |F_0(\tau_t)|^2,
\end{align}
where $\Sigma_{\rm NNLL}^{(2)}(p_{\rm t, veto})|_{{\rm large}-m_t}$ is
obtained from the second-order expansion of the large-$m_t$
resummation formula of ref.~\cite{Banfi:2012jm}.  The vetoed
cross-section, in the first of the three matching schemes, reads
\begin{multline}
\label{eq:scheme-a}
\Sigma^{\rm (a)}_{\rm matched}(p_{\rm t, veto}) =\frac{1}{\sigma_0}\frac{\Sigma_{\rm NNLL}(p_{\rm t, veto})}{1+{\mathcal L}^{(1)}(\tilde{L})/{\mathcal L}^{(0)}(\tilde{L})}
 \bigg[\sigma_0\left(1+\frac{{\mathcal L}^{(1)}(\tilde{L})}{{\mathcal L}^{(0)}(\tilde{L})}\right)+\Sigma^{(1)}(p_{\rm t, veto})
 -\Sigma_{\rm NNLL}^{(1)}(p_{\rm t, veto})\\
 +\Sigma^{(2)}(p_{\rm t, veto})-\tilde{\Sigma}_{\rm NNLL}^{(2)}(p_{\rm t, veto})
 +\left(\frac{{\mathcal L}^{(1)}(0)}{{\mathcal L}^{(0)}(0)}
 -\frac{\Sigma_{\rm NNLL}^{(1)}(p_{\rm t, veto})}{\sigma_{0}}\right)
 \left(\Sigma^{(1)}(p_{\rm t, veto})-\Sigma_{\rm NNLL}^{(1)}(p_{\rm t, veto})\right)\bigg]\,
\end{multline}
where 
\begin{equation}
\label{eq:Ltilde}
\tilde L=\frac{1}{p}\ln\left[1+\left(\frac{Q}{\ptjv}\right)^p\right]\,.
\end{equation}
For our numerical results we choose $p=5$~\cite{Banfi:2012yh}.

The luminosity factors ${\mathcal L}^{(0)}(\tilde{L})$ and ${\mathcal
  L}^{(1)}(\tilde{L})$ are defined as
\begin{align}
 \label{eq:L_0}
 {\mathcal L}^{(0)}(\tilde{L}) &= \sum_{i,j}\int dx_1 dx_2 \delta(x_1 x_2 s - M^2)
  f_i\!\left(x_1, e^{-\tilde{L}} \mu_F\right)f_j\!\left(x_2, e^{-\tilde{L}} \mu_F\right), \\ 
 {\mathcal L}^{(1)}(\tilde{L}) &=   \frac{\alpha_{s}}{2\pi}\sum_{i,j}\int dx_1 dx_2  \delta(x_1 x_2 s - M^2)
 \bigg[f_i\!\left(x_1, e^{-\tilde{L}} \mu_F\right)
  f_j\!\left(x_2, e^{-\tilde{L}} \mu_F\right){\cal H}^{(1)} \notag\\
  &+\frac{1}{1-2\alpha_s \beta_0 \tilde{L}}\sum_{k}\bigg(
  \int_{x_1}^1\frac{dz}{z} C_{ki}^{(1)}(z)
  f_i\!\left(\frac{x_1}{z}, e^{-\tilde{L}} \mu_F\right)
  f_j\!\left(x_2, e^{-\tilde{L}} \mu_F\right) + \{(x_1,i)\,\leftrightarrow\,(x_2,j)\}\bigg)\, \bigg].
\end{align}

The second scheme can be derived from the previous one by replacing
$\Sigma^{(2)}(p_{\rm t, veto})$ with  $\bar{\Sigma}^{(2)}(p_{\rm t, veto})$. For the 
vetoed cross section we get
\begin{multline}
\label{eq:scheme-b}
 \Sigma^{\rm (b)}_{\rm matched}(p_{\rm t, veto}) =\frac{1}{\sigma_0}\frac{\Sigma_{\rm NNLL}(p_{\rm t, veto})}{1+{\mathcal L}^{(1)}(\tilde{L})/{\mathcal L}^{(0)}(\tilde{L})}
 \bigg[\sigma_0\left(1+\frac{{\mathcal L}^{(1)}(\tilde{L})}{{\mathcal L}^{(0)}(\tilde{L})}\right)+\Sigma^{(1)}(p_{\rm t, veto})
 -\Sigma_{\rm NNLL}^{(1)}(p_{\rm t, veto})\\
 +\bar{\Sigma}^{(2)}(p_{\rm t, veto})-\tilde{\Sigma}_{\rm NNLL}^{(2)}(p_{\rm t, veto})
 +\left(\frac{{\mathcal L}^{(1)}(0)}{{\mathcal L}^{(0)}(0)}
 -\frac{\Sigma_{\rm NNLL}^{(1)}(p_{\rm t, veto})}{\sigma_{0}}\right)
 \left(\Sigma^{(1)}(p_{\rm t, veto})-\Sigma_{\rm NNLL}^{(1)}(p_{\rm t, veto})\right)\bigg].
\end{multline}
Finally, the third matching scheme is directly formulated for the efficiency resulting in 
\begin{multline}
\label{eq:scheme-c}
 \epsilon^{\rm (c)}_{\rm matched}(p_{\rm t, veto}) =\frac{1}{\sigma_0^2}\frac{\Sigma_{\rm NNLL}(p_{\rm t, veto})}{1+{\mathcal L}^{(1)}(\tilde{L})/{\mathcal L}^{(0)}(\tilde{L})} 
 \bigg[\sigma_0\left(1+\frac{{\mathcal L}^{(1)}(\tilde{L})}{{\mathcal L}^{(0)}(\tilde{L})}\right)
  +\bar{\Sigma}^{(1)}(p_{\rm t, veto})
 -\Sigma_{\rm NNLL}^{(1)}(p_{\rm t, veto})\\
 +\bar{\Sigma}^{(2)}(p_{\rm t, veto})-\frac{\sigma_1}{\sigma_0}\bar{\Sigma}^{(1)}(p_{\rm t, veto})-\tilde{\Sigma}_{\rm NNLL}^{(2)}(p_{\rm t, veto})\\
 +\left(\frac{{\mathcal L}^{(1)}(0)}{{\mathcal L}^{(0)}(0)}
 -\frac{\Sigma_{\rm NNLL}^{(1)}(p_{\rm t, veto})}{\sigma_{0}}\right)
 \left(\bar{\Sigma}^{(1)}(p_{\rm t, veto})-\Sigma_{\rm NNLL}^{(1)}(p_{\rm t, veto})\right)\bigg].
\end{multline}

The logR-(a) scheme discussed in the text to estimate the size of subleading mass effects reads
\begin{multline}
\label{eq:scheme-logR}
\Sigma^{\rm logR-(a)}_{\rm matched}(p_{\rm t, veto}) =\frac{1}{\sigma_0}\frac{\Sigma_{\rm NNLL}(p_{\rm t, veto})}{1+{\mathcal L}^{(1)}(\tilde{L})/{\mathcal L}^{(0)}(\tilde{L})}
 \bigg[\sigma_0\left(1+\frac{{\mathcal L}^{(1)}(\tilde{L})}{{\mathcal L}^{(0)}(\tilde{L})}\right)
 +\Sigma^{(2)}(p_{\rm t, veto})-\tilde{\Sigma}_{\rm NNLL}^{(2)}(p_{\rm t, veto})\\
 -\frac{\Sigma_{\rm NNLL}^{(1)}(p_{\rm t, veto})}{\sigma_{0}}
 \left(\Sigma^{(1)}(p_{\rm t, veto})-\Sigma_{\rm NNLL}^{(1)}(p_{\rm t, veto})\right)\bigg]\\
\exp\bigg\{\frac{\Sigma^{(1)}(p_{\rm t, veto})-\Sigma_{\rm NNLL}^{(1)}(p_{\rm t, veto})}{\sigma_0}
-\frac{1}{2}\left(\frac{\Sigma^{(1)}(p_{\rm t, veto})-\Sigma_{\rm NNLL}^{(1)}(p_{\rm t, veto})}{\sigma_0}
 \right)^2\bigg\}\,.
\end{multline}
Notice that the (c) scheme is directly expressed in terms of the jet
veto efficiency, while for the schemes (a), (b), logR-(a) and mod-(a)~eq.~\eqref{eq:moda-scheme} the
corresponding efficiencies are defined as
\begin{equation}
 \epsilon_{\rm matched}(p_{\rm t, veto}) = \frac{\Sigma_{\rm matched}(p_{\rm t, veto})}{\Sigma_{\rm matched}(p_{\rm t, veto}^{\rm max})}\,,
\end{equation}
where $\ptjv^{\rm max}$ is maximum kinematically allowed jet transverse momentum.

\section{Mod-(a) matching scheme}
\label{sec:modascheme}

In the matching schemes of ref.~\cite{Banfi:2012jm} used here, for
$\ptjv \ll m_H$ the remainder is multiplied by both a Sudakov form
factor and a luminosity prefactor (see
eqs.~\eqref{eq:scheme-a},~\eqref{eq:scheme-b},~\eqref{eq:scheme-c}).
It is actually not known whether this structure is fulfilled by higher
order corrections. One might argue that multiplying the remainder by a
Sudakov form factor is appropriate, since the latter represents the
probability of having no emissions at scales above $\ptjv$. On the
other hand, the luminosity prefactor accounts for all collinear
splittings which happen at scales smaller than $\ptjv$. This is the
same transverse momentum region which contributes to the remainder
function. Therefore, it might seem more appropriate to devise a new
matching scheme where the remainder is not multiplied neither by the 
Sudakov form factor nor by a luminosity prefactor.

The new hybrid scheme, referred to as mod-(a) scheme, is a
modification of the (a) scheme of ref.~\cite{Banfi:2012jm}, where the
${\cal O}(\alpha_s^3)$ regular remainder is treated additively. We
hence introduce the mod-(a) scheme as

\begin{multline}
\label{eq:moda-scheme}
\Sigma^{\rm mod-(a)}_{\rm matched}(p_{\rm t, veto}) =\frac{1}{\sigma_0}\frac{\Sigma_{\rm NNLL}(p_{\rm t, veto})}{1+{\mathcal L}^{(1)}(\tilde{L})/{\mathcal L}^{(0)}(\tilde{L})}
 \bigg[\sigma_0\left(1+\frac{{\mathcal L}^{(1)}(\tilde{L})}{{\mathcal L}^{(0)}(\tilde{L})}\right)
 +\Sigma^{(2)}(p_{\rm t, veto})-\tilde{\Sigma}_{\rm NNLL}^{(2)}(p_{\rm t, veto})\bigg]\\
+\Sigma^{(1)}(p_{\rm t, veto}) - \Sigma_{\rm NNLL}^{(1)}(p_{\rm t, veto}).
\end{multline}
This scheme is defined as scheme (a), but the one-loop remainder terms
are not multiplied by the luminosity or Sudakov form factors. 
Instead, the two-loop remainder is treated in a multiplicative way, as
in the original formulation of the (a) scheme. In this way, the
matched cross-section contains all logarithmically enhanced
contributions in $\Sigma^{(J)}(\ptjv)$ up to order $\alpha_s^n
L^{2n-3}$. 
We remind however that the fixed-order results that we use do not
contain the correct ${\cal O}(\alpha_s^4)$ constant term with full
dependence on the heavy-quark masses. Therefore the resulting two-loop
constant terms is rescaled according to the correction factor $|
F_0(\tau_t)|^2$ where $F_0(\tau_t)$ is defined in~\eqref{eq:f0tau}.

\begin{figure}[htp]
  \centering
  \includegraphics[width=0.70\columnwidth]{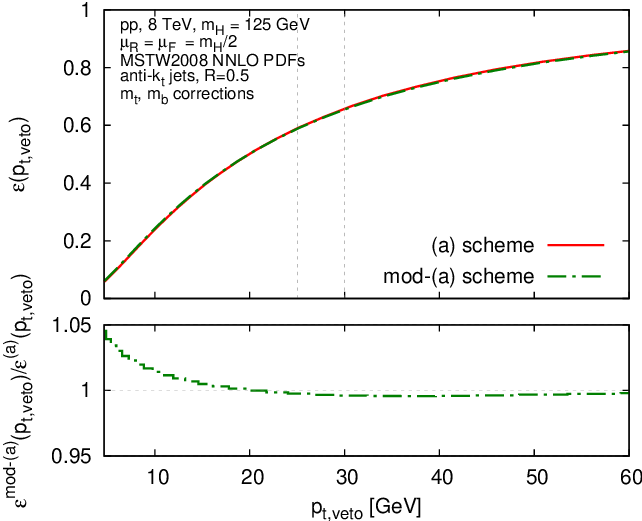}
  \caption{Matched jet veto efficiency obtained with two different
    matching schemes. The two schemes differ in the treatment of the
    NLO bottom-mass-dependent remainder as discussed in the text.}
  \label{fig:matching-schemes}
\end{figure}
Figure~\ref{fig:matching-schemes} shows the comparison between the
mod-(a) scheme and the (a) scheme for the matched jet veto efficiency.
The difference between the two schemes gives us an estimate of whether
it is numerically relevant to multiply the remainder by the Sudakov
and luminosity factors, or not.  We see that the two results differ at
the few-percent level for very small $\ptjv \lesssim 10$ GeV, but only
at per-mille level in the region of interest $\ptjv \sim 25$-$30$ GeV.

\section{Comparison to the Higgs transverse momentum}
\label{app:ptjvspth}

It is interesting to investigate whether, in our framework,
mass-effects affect the Higgs transverse momentum distribution in a
similar way to the jet-veto distribution. We have therefore extended the
\textsc{JetVHeto} code to produce resummed and matched results for the
distribution in the Higgs transverse momentum $p_{t,H}$, including
mass effects. For this purpose, we have used eq.~(43) of
ref.~\cite{Banfi:2012jm}, with the same one-loop coefficient constants
as for jet-veto distribution (${\cal H}^{(1)}$ and $C_{ki}^{(1)}(z)$
in eq.~(\ref{eq:SigmaNNLL-result})).

We first note that, already at NLL, the multiple-emission function for $p_{t,H}$, ${\cal F}
(R')= e^{-\gamma_E R'}\Gamma(1-R'/2)/\Gamma(1+R'/2)$, has a divergence
at $R'=2$, which corresponds to $p_{t,H}\sim 5$ GeV for a resummation
scale $Q=m_H/2$. Decreasing the value of $Q$ pushes the divergence to
even lower values of $p_{t,H}$. The origin of this divergence is well
understood~\cite{Frixione:1998dw} and is explained in detail in the
case of the oblateness in ref.~\cite{Banfi:2001bz}. It is related to
the fact that multiple emissions contribute vectorially to $p_{t,H}$,
therefore a small value of $p_{t,H}$ can also arise in configurations
involving emissions with large $p_t$ that cancel against each
other. When this mechanism dominates over the standard Sudakov
suppression (events for which the low $p_{t,H}$ is due to low $p_t$
emissions), one can not use the normal logarithmic hierarchy to
establish which logarithmic terms are dominant.
The divergence in $\cF(R')$ is related to using a formula that assumes
this hierarchy, obtained by neglecting subleading effects. Since our
prediction diverges at $p_{t,H} \sim 5$ GeV, we consider it unreliable
for $p_{t,H} \lesssim 15$ GeV.  The excluded region is denoted by a
gray band in the plots below.

\begin{figure}[htp]
  \centering
  \includegraphics[width=1.05\columnwidth]{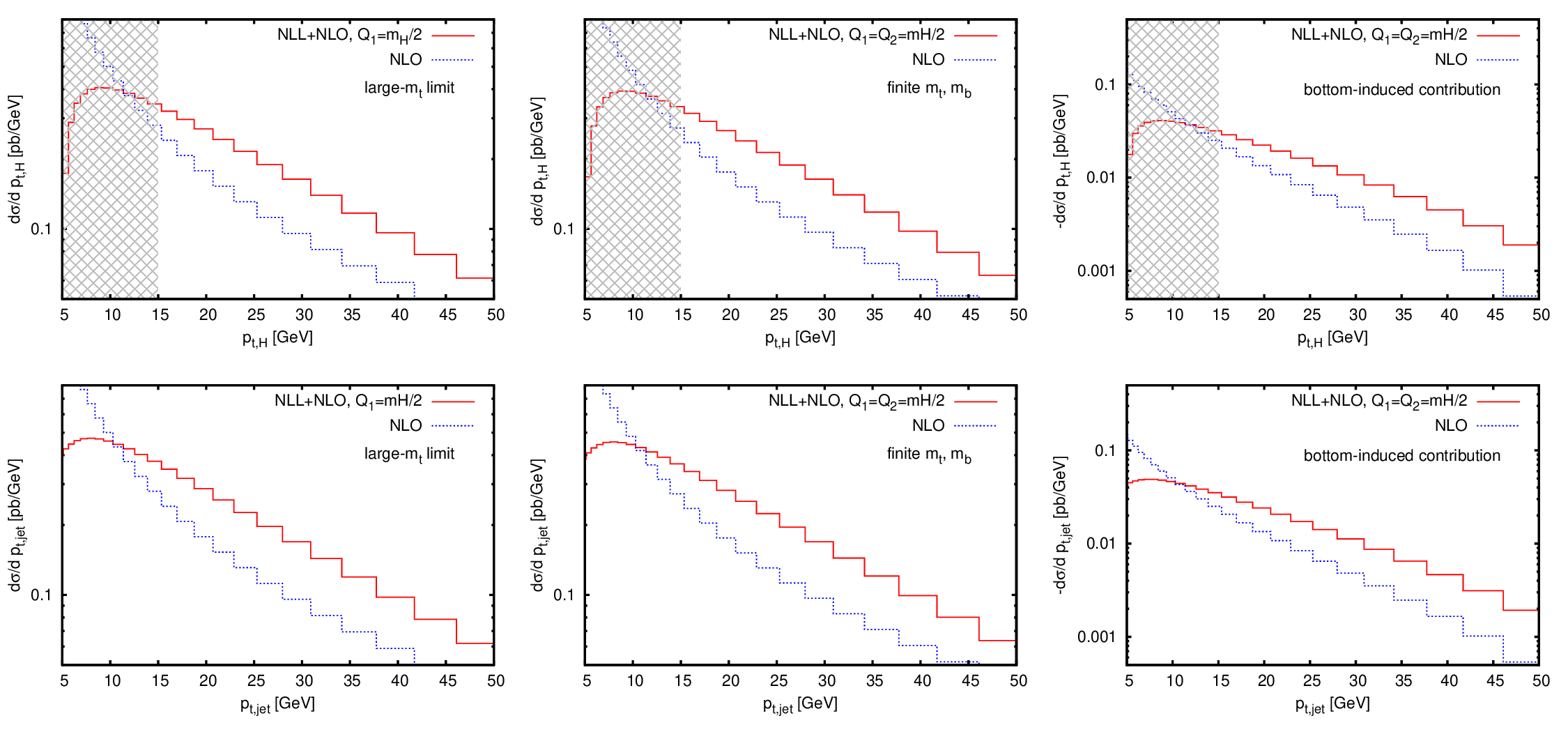}
  \caption{NLL+NLO (red) and NLO (blue) differential distributions for
    $p_{t,H}$ (upper plots) and $p_{\rm t, jet}$ (lower plots) in the
    large-$m_t$ limit (left-hand plots), including finite $m_t,
    m_b$ effects (central plots) and bottom-induced contribution only (right-hand plots).}
  \label{fig:ptdist-nlo}
\end{figure}

\begin{figure}[htp]
  \centering
  \includegraphics[width=1.05\columnwidth]{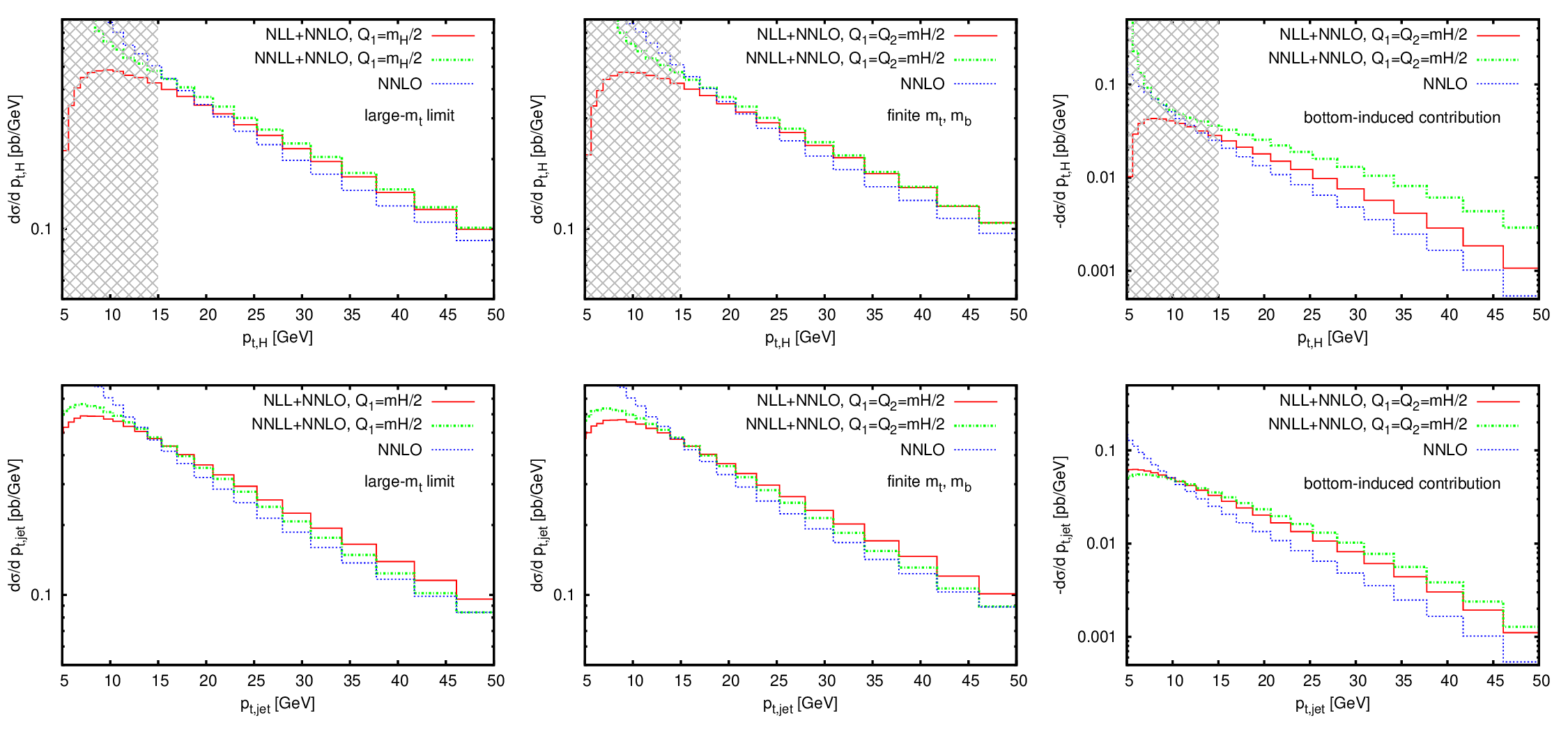}
  \caption{NLL+NNLO (red), NNLL+NNLO (green) and NNLO (blue)
    differential distributions for $p_{t,H}$ (upper plots) and $p_{\rm
      t, jet}$ (lower plots) in the large-$m_t$ limit (left-hand
    plots), including finite $m_t, m_b$ effects (central plots) and
    bottom-induced contribution only (right-hand plots)}
  \label{fig:ptdist-nnlo}
\end{figure}

In the following, we wish to investigate the origin of the large
discrepancy between matched and fixed-order distributions observed by
the authors of ref.~\cite{Grazzini:2013mca}, when using a resummation
scale of order $m_H$.  We then consider matched distributions obtained
with $\mu_R=\mu_F=m_H$ and resummation scales $Q_1=Q_2=m_H/2$ (where
$Q_1$ and $Q_2$ are the resummation scales corresponding to the top
and bottom-induced contributions, respectively).
More specifically, in Figs.~\ref{fig:ptdist-nlo}
(\ref{fig:ptdist-nnlo}) we show the NLL+NLO and NLO (NLL+NNLO,
NNLL+NNLO and NNLO) differential distributions for $p_{t,H}$ (upper
plots) and $p_{\rm t, jet}$ (lower plots) in the large-$m_t$ limit
(left-hand plots), and including finite $m_t, m_b$ effects (central
plots). Since the bottom-induced contribution is only a small fraction
of the total, it is useful to also plot this separately (right-hand
plots).
We observe that all of the NLL+NLO matched results differ
considerably from the NLO in the intermediate $p_t$ region ($20\,
\mathrm{ GeV} \lesssim p_t \lesssim 50\, \mathrm{ GeV}$).  This
happens regardless of the observable, $p_{t,H}$ or $p_{\rm t, jet}$,
and of whether one includes mass-effects, or not.
We therefore conclude that this discrepancy cannot be ascribed to a
large non-factorizing correction, which is present only in the
bottom-induced contributions. The difference between the matched and
fixed-order results is due to missing higher order terms.

In the large-$m_t$ limit (and in all standard cases), when a matching
is performed at the NLL+NLO (NNLL+NNLO) level, this difference is of
relative order $\cO{\as^2}$ ($\cO{\as^3}$ ). Therefore we do observe a
better convergence with a NNLL+NNLO matching, as one would expect (see
left-hand plots of Fig.~\ref{fig:ptdist-nnlo}).
Unfortunately, the full NNLO correction ($\cO{\as^3}$) including exact
mass effects is so far unknown. Therefore we use the NNLO result in
the large $m_t$-limit, rescaled by the ratio of the Born cross-section
including exact top-mass effects to the Born cross-section in the
large $m_t$ limit. When matching to this NNLO correction, the
expansion of our resummation formula at $\cO{\as^2}$ appearing in our
matching formulae is replaced by the corresponding one in the large $m_t$ limit
multiplied by the same rescaling factor. This is to guarantee NNLL
accuracy in the Sudakov region. However, this implies that even when matching
at NNLO+NNLL level, the discrepancy between fixed-order and matched
results is formally still $\cO{\as^2}$.

\begin{figure}[htp]
  \centering
  \includegraphics[width=1.0\columnwidth]{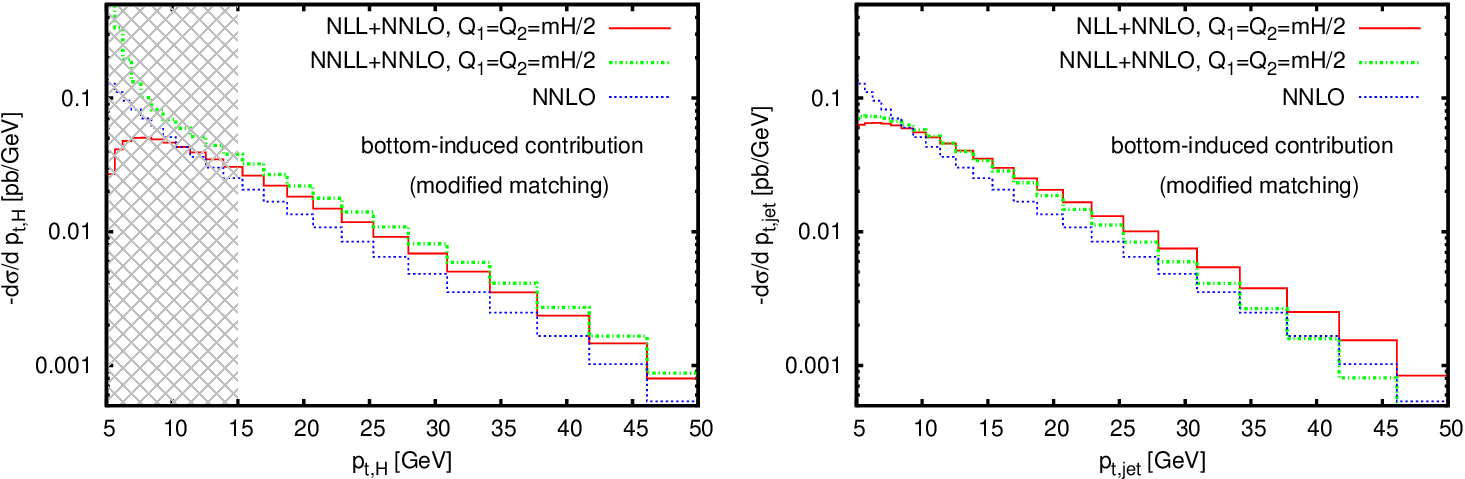}
  \caption{
NLL+NNLO (red), NNLL+NNLO (green) and NNLO (blue) differential
distributions for the bottom-induced contribution for $p_{t,H}$ (left)
and $p_{\rm t, jet}$ (right) using the actual expansion of the resummation
formula when matching to NNLO (see text for more details).
  \label{fig:wrong}}
\end{figure}

In order to investigate whether this is really the source of the
observed difference between matched and fixed-order distributions, we
can use the actual expansion of our resummation (which however spoils
the logarithmic accuracy in the low $p_t$ region). This ensures that,
in the medium-large $p_t$ region, the correct fixed-order behavior is
reproduced, and the left-over is of $\cO{\as^3}$.  The outcome of this
exercise is shown in Fig.~\ref{fig:wrong} for the bottom-induced
contribution. As expected, the discrepancy between matched and
fixed-order is now similar to the one observed in the large $m_t$ limit.

\begin{thebibliography}{99}


\bibitem{Aad:2012tfa}
  G.~Aad {\it et al.}  [ATLAS Collaboration],
  Phys.\ Lett.\ B {\bf 716} (2012) 1
  [arXiv:1207.7214 [hep-ex]].

\bibitem{Chatrchyan:2012ufa}
  S.~Chatrchyan {\it et al.}  [CMS Collaboration],
  Phys.\ Lett.\ B {\bf 716} (2012) 30
  [arXiv:1207.7235 [hep-ex]].


\bibitem{Banfi:2012yh}
  A.~Banfi, G.~P.~Salam and G.~Zanderighi,
  JHEP {\bf 1206} (2012) 159
  [arXiv:1203.5773 [hep-ph]].

\bibitem{Banfi:2004yd}
  A.~Banfi, G.~P.~Salam and G.~Zanderighi,
  JHEP {\bf 0503} (2005) 073
  [hep-ph/0407286].

\bibitem{Becher:2012qa}
  T.~Becher and M.~Neubert,
  JHEP {\bf 1207} (2012) 108
  [arXiv:1205.3806 [hep-ph]].

\bibitem{Banfi:2012jm}
  A.~Banfi, P.~F.~Monni, G.~P.~Salam and G.~Zanderighi,
  Phys.\ Rev.\ Lett.\  {\bf 109} (2012) 202001
  [arXiv:1206.4998 [hep-ph]].

\bibitem{Stewart:2013faa}
  I.~W.~Stewart, F.~J.~Tackmann, J.~R.~Walsh and S.~Zuberi,
  arXiv:1307.1808 [hep-ph].

\bibitem{Becher:2013xia}
  T.~Becher, M.~Neubert and L.~Rothen,
  arXiv:1307.0025 [hep-ph].

\bibitem{Cacciari:2008gp}
  M.~Cacciari, G.~P.~Salam, G.~Soyez,
  JHEP {\bf 0804 } (2008)  063
  [arXiv:0802.1189 [hep-ph]].



\bibitem{Spira:1995rr}
  M.~Spira, A.~Djouadi, D.~Graudenz and P.~M.~Zerwas,
  Nucl.\ Phys.\ B {\bf 453} (1995) 17
  [hep-ph/9504378].

\bibitem{Spira:1997dg}
  M.~Spira,
  Fortsch.\ Phys.\  {\bf 46} (1998) 203
  [hep-ph/9705337].

\bibitem{Harlander:2005rq}
  R.~Harlander and P.~Kant,
  JHEP {\bf 0512} (2005) 015
  [hep-ph/0509189].


\bibitem{Anastasiou:2006hc}
  C.~Anastasiou, S.~Beerli, S.~Bucherer, A.~Daleo and Z.~Kunszt,
  JHEP {\bf 0701} (2007) 082
  [hep-ph/0611236].


\bibitem{Aglietti:2006tp}
  U.~Aglietti, R.~Bonciani, G.~Degrassi and A.~Vicini,
  JHEP {\bf 0701} (2007) 021
  [hep-ph/0611266].
\bibitem{Bonciani:2007ex}
  R.~Bonciani, G.~Degrassi and A.~Vicini,
  JHEP {\bf 0711} (2007) 095
  [arXiv:0709.4227 [hep-ph]].

\bibitem{Harlander:2012hf}
  R.~V.~Harlander, T.~Neumann, K.~J.~Ozeren and M.~Wiesemann,
  JHEP {\bf 1208} (2012) 139
  [arXiv:1206.0157 [hep-ph]].


\bibitem{MCFM}
http://mcfm.fnal.gov/

\bibitem{Anastasiou:2009kn}
  C.~Anastasiou, S.~Bucherer and Z.~Kunszt,
  JHEP {\bf 0910} (2009) 068
  [arXiv:0907.2362 [hep-ph]].

\bibitem{Harlander:2012pb}
  R.~V.~Harlander, S.~Liebler and H.~Mantler,
  Computer Physics Communications {\bf 184} (2013) pp. 1605
  [arXiv:1212.3249 [hep-ph]].

\bibitem{Harlander:2003ai}
  R.~V.~Harlander and W.~B.~Kilgore,
  Phys.\ Rev.\ D {\bf 68} (2003) 013001
  [hep-ph/0304035].

\bibitem{Harlander:2010cz}
  R.~V.~Harlander, K.~J.~Ozeren and M.~Wiesemann,
  Phys.\ Lett.\ B {\bf 693} (2010) 269
  [arXiv:1007.5411 [hep-ph]].

\bibitem{Harlander:2011fx}
  R.~Harlander and M.~Wiesemann,
  JHEP {\bf 1204} (2012) 066
  [arXiv:1111.2182 [hep-ph]].

\bibitem{Buehler:2012cu}
  S.~Buehler, F.~Herzog, A.~Lazopoulos and R.~Mueller,
  JHEP {\bf 1207} (2012) 115
  [arXiv:1204.4415 [hep-ph]].


\bibitem{Corcella:2000bw}
  G.~Corcella, I.~G.~Knowles, G.~Marchesini, S.~Moretti, K.~Odagiri, P.~Richardson, M.~H.~Seymour and B.~R.~Webber,
  JHEP {\bf 0101} (2001) 010
  [hep-ph/0011363].

\bibitem{Corcella:2002jc} 
  G.~Corcella, I.~G.~Knowles, G.~Marchesini, S.~Moretti, K.~Odagiri, P.~Richardson, M.~H.~Seymour and B.~R.~Webber,
  hep-ph/0210213.


\bibitem{Sjostrand:2006za}
  T.~Sjostrand, S.~Mrenna and P.~Z.~Skands,
  JHEP {\bf 0605} (2006) 026
  [hep-ph/0603175].


\bibitem{Bagnaschi:2011tu}
  E.~Bagnaschi, G.~Degrassi, P.~Slavich and A.~Vicini,
  JHEP {\bf 1202} (2012) 088
  [arXiv:1111.2854 [hep-ph]].

\bibitem{mcatnlo} From Version 4.08, see
  http://www.hep.phy.cam.ac.uk/theory/webber/MCatNLO/

\bibitem{Mantler:2012bj}
  H.~Mantler and M.~Wiesemann,
  arXiv:1210.8263 [hep-ph].


\bibitem{Grazzini:2013mca}
  M.~Grazzini and H.~Sargsyan,
  arXiv:1306.4581 [hep-ph].

\bibitem{Ellis:2011cr}
  R.~K.~Ellis, Z.~Kunszt, K.~Melnikov and G.~Zanderighi,
  Phys.\ Rept.\  {\bf 518} (2012) 141
  [arXiv:1105.4319 [hep-ph]].

\bibitem{Ellis:1987xu}
  R.~K.~Ellis, I.~Hinchliffe, M.~Soldate and J.~J.~van der Bij,
  Nucl.\ Phys.\ B {\bf 297} (1988) 221.

\bibitem{Baur:1989cm}
  U.~Baur and E.~W.~N.~Glover,
  Nucl.\ Phys.\ B {\bf 339} (1990) 38.



\bibitem{Catani:1992ua}
  S.~Catani, L.~Trentadue, G.~Turnock and B.~R.~Webber,
  Nucl.\ Phys.\ B {\bf 407} (1993) 3.
 

\bibitem{Martin:2009iq}
  A.~D.~Martin, W.~J.~Stirling, R.~S.~Thorne and G.~Watt,
  Eur.\ Phys.\ J.\ C {\bf 63} (2009) 189
  [arXiv:0901.0002 [hep-ph]].
  
\bibitem{FastJet} 
  M.~Cacciari and G.~P.~Salam,
  Phys.\ Lett.\  B {\bf 641} (2006) 57
  [arXiv:hep-ph/0512210];\\
  M.~Cacciari, G.~P.~Salam and G.~Soyez,
  arXiv:1111.6097 [hep-ph].


\bibitem{Catani:2007vq}
  S.~Catani and M.~Grazzini,
  Phys.\ Rev.\ Lett.\  {\bf 98} (2007) 222002
  [hep-ph/0703012].

\bibitem{Heinemeyer:2013tqa}
  S.~Heinemeyer {\it et al.}  [ The LHC Higgs Cross Section Working Group Collaboration],
  arXiv:1307.1347 [hep-ph].



\bibitem{Nason:2004rx}
  P.~Nason,
  JHEP {\bf 0411} (2004) 040
  [hep-ph/0409146].

\bibitem{Alioli:2010xd}
  S.~Alioli, P.~Nason, C.~Oleari and E.~Re,
  JHEP {\bf 1006} (2010) 043
  [arXiv:1002.2581 [hep-ph]].

\bibitem{Frixione:2002ik}
  S.~Frixione and B.~R.~Webber,
  JHEP {\bf 0206} (2002) 029
  [hep-ph/0204244].

\bibitem{Hamilton:2012np}
  K.~Hamilton, P.~Nason and G.~Zanderighi,
  JHEP {\bf 1210} (2012) 155
  [arXiv:1206.3572 [hep-ph]].

\bibitem{Hamilton:2012rf}
  K.~Hamilton, P.~Nason, C.~Oleari and G.~Zanderighi,
  JHEP {\bf 1305} (2013) 082
  [arXiv:1212.4504 [hep-ph]].

\bibitem{vicinitalk}
Talk given by A. Vicini,
http://indico.cern.ch/getFile.py/access?contribId=2\&resId=0\&materialId=slides\&confId=263472. 

\bibitem{jetvheto}
http://jetvheto.hepforge.org/

\bibitem{Frixione:1998dw}
  S.~Frixione, P.~Nason and G.~Ridolfi,
  Nucl.\ Phys.\ B {\bf 542} (1999) 311
  [hep-ph/9809367].

\bibitem{Banfi:2001bz}
  A.~Banfi, G.~P.~Salam and G.~Zanderighi,
  JHEP {\bf 0201} (2002) 018
  [hep-ph/0112156].


\end{thebibliography}
\end{document}